\documentclass{aa} 
\usepackage{natbib}
\usepackage{graphicx}
\usepackage{txfonts}

\begin{document}

\title{ The $0.4<z<1.3$ star formation history of the Universe as viewed in the far-infrared}
\author{B. Magnelli\inst{1}
\and
D. Elbaz\inst{1}
\and
R.~R Chary\inst{2}
\and
M. Dickinson\inst{3}
\and
D. Le Borgne\inst{1,4}
\and
D. T. Frayer\inst{2} 
\and
C. N. A. Willmer\inst{5}}
\institute{Laboratoire AIM, CEA/DSM-CNRS-Universit\'e Paris Diderot, IRFU/Service d'Astrophysique, B\^at. 709, CEA-Saclay, F-91191 Gif-sur-Yvette C\'edex, France\\
\email{Benjamin.magnelli@cea.fr}
\and
Spitzer Science Center, California Institute of Technology, Pasadena, CA 91125, USA
\and
National Optical Astronomy Observatory, Tucson, AZ 85719, USA
\and
Institut d'Astrophysique de Paris, UMR7095 CNRS, UPMC, 98bis boulevard Arago, F-75014 Paris, France
\and
Steward Observatory, University of Arizona, 933 North Cherry Avenue, Tucson, AZ 85721, USA}
\date{Received ??; accepted ??}

\abstract
{}
{
We use the deepest existing mid- and far-infrared observations (reaching $\sim$3 mJy at 70\,$\mu$m) obtained with \textit{Spitzer} in the Great Observatories Origins Deep Survey (GOODS) and Far Infrared Deep Extragalactic Legacy survey (FIDEL) fields to derive the evolution of the rest-frame 15 $\mu$m, 35\,$\mu$m, and total infrared luminosity functions of galaxies spanning $z<1.3$. 
We thereby quantify the fractional contribution of infrared luminous galaxies to the comoving star formation rate density over this redshift range.
In comparison with previous studies, the present one takes advantage of deep 70 $\mu$m observations that provide a more robust infrared luminosity indicator than 24 $\mu$m affected by the emission of PAHs at high redshift ($z\thicksim1$), and we use several independent fields to control cosmic variance.
}
{
We used a new extraction technique based on the well-determined positions of galaxies at shorter wavelengths to extract the 24 and 70\,$\mu$m flux densities of galaxies.
It is found that sources separated by a minimum of 0.5$\times$FWHM are deblended by this technique, which facilitates multi-wavelength associations of counterparts.
Using a combination of photometric and spectroscopic redshifts that exist for $\sim$80\% of the sources in our sample, we are able to estimate the rest-frame luminosities of galaxies at 15\,$\mu$m and 35\,$\mu$m.
By complementing direct detections with a careful stacking analysis, we measured the mid- and far-infrared luminosity functions of galaxies over a factor $\sim$100 in luminosity ($\rm{10^{11}\,L_{\odot}<\thicksim L_{IR}<\thicksim 10^{13}\,L_{\odot}}$) at $z<1.3$.
A stacking analysis was performed to validate the bolometric corrections and to compute comoving star-formation rate densities in three redshift bins $0.4<z<0.7$, $0.7<z<1.0$ and, $1.0<z<1.3$.
}
{
We find that the average infrared spectral energy distribution of galaxies over the last 2/3 of the cosmic time is consistent with that of local galaxies, although individual sources do present significant scatter.
We also measured both the bright and faint ends of the infrared luminosity functions and find no evidence for a change in the slope of the double power law used to characterize the luminosity function.
The redshift evolution of infrared luminous galaxies is consistent with pure luminosity evolution  proportional to $(1+z)^{3.6\pm0.4}$ up to $z\thicksim1.3$.
We do not find evidence of differential evolution between LIRGs and ULIRGs up to $z\thicksim1.3$, in contrast with previous claims. The comoving number density of infrared luminous galaxies has increased by a factor of $\thicksim100$ between $0<z<1$.
By $z\sim1.0$, LIRGs produce half of the total comoving infrared luminosity density.
}
{}
\authorrunning{Magnelli et al. }
\titlerunning{Star Formation Rate Density at $ 0<z<1.3$}
\maketitle

\section{Introduction}
\indent{
Constraining the star formation history of galaxies as a function of redshift is a key to understanding galaxy formation.
To estimate the ongoing star formation rate (SFR) of galaxies one needs to measure direct and re-radiated stellar emission of young stars which correspond to the rest-frame UV and IR light, respectively. 
Due to the difficulties and the only recent possibility to obtain deep infrared observations, first studies which have estimated SFR evolution with redshift have only used UV observations corrected for dust extinction \citep{Madau1999}.
However dust extinction correction suffers from various uncertainties and it is well known that local galaxies harboring strong dusty star formation are opaque to UV radiation (Buat et al. 2005).
Hence to obtain a complete understanding of these dusty star forming galaxies, deep infrared observations are required.
\\}
\indent{
Taking advantage of infrared capabilities such as the \textit{Infrared Space Observatory} (ISO), several teams have confirmed the importance of infrared galaxies for the understanding of SFR evolution \citep{elbaz_1999,aussel_1999, chary_2001, franceschini_2001, xu_2001, elbaz_2002, metcalfe_2003, lagache_2004}.
Indeed they all found that the contribution of luminous infrared galaxies to the SFR density increases with redshift up to $z\sim$ 1 and that at high redshift the bulk of the SFR density occurs in luminous infrared galaxies (i.e $10^{11} L_{\odot}\leq\,$LIRGs\,$ <10^{12} L_{\odot},\ 10^{12} L_{\odot}\leq\,$ULIRGs) which are relatively rare in the local Universe.  
\\}
\indent{
The Spitzer Space Telescope made it possible to obtain a complete census of the SFR density evolution up to $z\thicksim1$ using deep 24 $\mu$m Multiband Imaging Photometer \citep[MIPS;][]{rieke_2004} observations.
Many authors \citep{lefloch_2005, chary_2001,lagache_2003,xu_2000} have studied the evolution of the rest-frame 15 $\mu$m or 8 $\mu$m luminosity function (LF) based on MIPS 24 $\mu$m data.
Then using empirical SED libraries \citep{chary_2001,lagache_2003,dale_2002} to derive the total infrared luminosities of galaxies from these monochromatic luminosities, they were able to constrain the evolution of the total infrared LF.
They all found a strong evolution with cosmic time of the relative contributions of normal, LIRG and ULIRG galaxies to the SFR density.
At $z\thicksim0$, the SFR density is dominated by quiescent galaxies whereas at $z\thicksim1$ it is dominated by LIRGs.
They also found that the contribution of ULIRGs to the SFR density rises steeply from $z=0$ to $z=1$ but still remains negligible at $z\thicksim1$ (i.e. less than 10\%).
More recently, studies using MIPS 24 $\mu$m \citep{caputi_2007} data have extended previous works to higher redshift and found that ULIRGs may dominate the SFR density at $z\thicksim2$.
\\}
\indent{
All these studies, based on 24 $\mu$m observations, are strongly dependent on the SED library \citep{chary_2001,lagache_2003,dale_2002} used to derive the total infrared luminosity of galaxies.
Moreover significant uncertainties on the bolometric correction are implied by the redshifting of the strong PAH and silicate emission into the 24 $\mu$m bandwidth.
\\}
\indent{
At $z\thicksim0$, \citet{calzetti_2007} have proven that the rest-frame 24 $\mu$m luminosity is a robust SFR estimator.
Moreover using deep radio observations \citet{appleton_2004} have proven that observed 70 $\mu$m flux was a more robust SFR estimator than the observed 24 $\mu$m flux.
Hence using the observed MIPS 70 $\mu$m data for galaxies below $z\sim$ 1.3, as in the present study, will strongly reduce the bolometric correction uncertainties since this emission corresponds to the rest-frame 35 $\mu$m luminosity at $z\thicksim1$ which is produced by dust of about the same temperature ($\rm{T\thicksim40\,K}$) as that responsible for the 24 $\mu$m emission in local galaxies.
The aim of the present paper is to take advantage of the deepest 70 $\mu$m observations currently available to strongly constrain the bright end slope of the 35 $\mu$m and the total infrared LFs up to $z\thicksim1.3$.
These far-infrared data, as part of the Far Infrared Deep Extragalactic Survey (FIDEL) and the Great Observatories Origin Deep Survey (GOODS), cover 4 fields (Extended Groth Strip, Extended Chandra Deep Field South, GOODS-N and, GOODS-S) which are wide enough ($\thicksim 1400$ total arcmin$^{2}$) and deep enough ($\thicksim 3 $mJy at 70 $\mu$m) to obtain a good constraint on the bright end slope of these LFs.
By complementing direct 70 $\mu$m detections with a careful stacking analysis, we also constrain the faint end slope of these LFs.
Finally, by studying the redshift evolution of the total infrared LF, we derive the evolution of the SFR density and the relative contributions of normal, LIRG and ULIRG galaxies to the total comoving infrared luminosity density as a function of redshift.
\\ \\}
\indent{
The layout of the paper is as follows:
In Section \ref{sec:data}, we present the mid- and far-infrared observations as well as all the ancillary data used in this study.
We also present in this section the new extraction technique used to extract infrared sources and the simulations made to characterize the limits of our infrared catalogs.
In Section \ref{sec:15 LF}, we present the rest-frame 15 $\mu$m LF inferred in three redshift bins center at $z=0.55,\ 0.85$ and, $1.15$ using MIPS 24 $\mu$m data.
The rest-frame 35 $\mu$m LF inferred in these three redshift bins using the MIPS 70 $\mu$m data is presented in the Section \ref{section: function 35}.
In this Section we also measure, using a stacking analysis, the observed 24/70 $\mu$m correlation which enables us to constrain the faint end slope of the rest frame 35 $\mu$m LF. 
In Section \ref{sec:ir LF} we present the total infrared LF function and its evolution up to $z\thicksim1.3$.
Finally in Section \ref{sec: discussion} we discuss the evolution of the rest-frame 35 $\mu$m and total infrared LF in terms of the cosmic SFR history and the relative contributions of normal and luminous infrared galaxies.
\\ \\}
\indent{
Throughout this paper we will use a cosmology with $H_{0}=71\ km\ s^{-1}\ Mpc^{-1},\Omega_{\Lambda}=0.73 ,\Omega_{M}=0.27$.\\
}
\section{The Data}
\label{sec:data}
\subsection{Infrared Imaging}
\subsubsection{EGS}
\indent{
The Extended Groth Strip field ($14^{h}17^{m},\,+52^{\circ}30\arcmin$) was observed at 24 and 70 $\mu$m with the MIPS instrument on board the \textit{Spitzer Space Telescope} as part of the Far Infrared Deep Extragalactic Legacy program (FIDEL, PI: Dickinson).
These observations cover a total area of 900 arcmin$^{2}$ using the scan mode AOT of \textit{Spitzer} (see \textit{Spitzer} observer's manual for more information). \\
}
\indent{
The data were coadded using the MOPEX mosaicing software provided by the \textit{Spitzer Science Center}. 
The final 24 $\mu$m mosaic has a pixel scale of 1.2$\arcsec/$pixel and an effective integration time per sky pixel of $\thicksim14\,000$s.
The final 70 $\mu$m mosaic has a pixel scale of 4.0$\arcsec/$pixel and an effective integration time per sky pixel of $\thicksim7\,200$s.
For both wavelengths the calibration factor used to convert digital units ($\rm{DN/s}$) to flux ($MJy/sr$) was taken from the MIPS data handbook (see SSC MIPS Data Handbook v3.3.1 section 3.7.2). 
At these wavelengths the point spread functions (PSF) are characterized by a FWHM of $\thicksim5.9\arcsec$ and $\thicksim18\arcsec$ at 24 and 70 $\mu$m observations, respectively.
These FWHM were measured on the empirical PSF used to perform our PSF fitting analysis (see Section \ref{subsec: prior method}).\\
}
\indent{
The 24 and 70 $\mu$m observations reach a point source sensitivity of $\thicksim50\ \mu$Jy and $\thicksim3\,000\ \mu$Jy respectively.
These point source sensitivities include effects of background and confusion noise and were derived using extensive simulations (see Section \ref{subsec: prior method} for a complete definition).
For both wavelengths, source extraction was first performed with a PSF fitting technique using the positions of IRAC sources as priors (see Section \ref{subsec: prior method}).
This procedure leads to a total of 8457 sources at 24 $\mu$m ($\ge 50\,\mu$Jy) and 669 sources at 70 $\mu$m ($\ge 3$ mJy).
}
\begin{figure*}
	\includegraphics[width=9.cm]{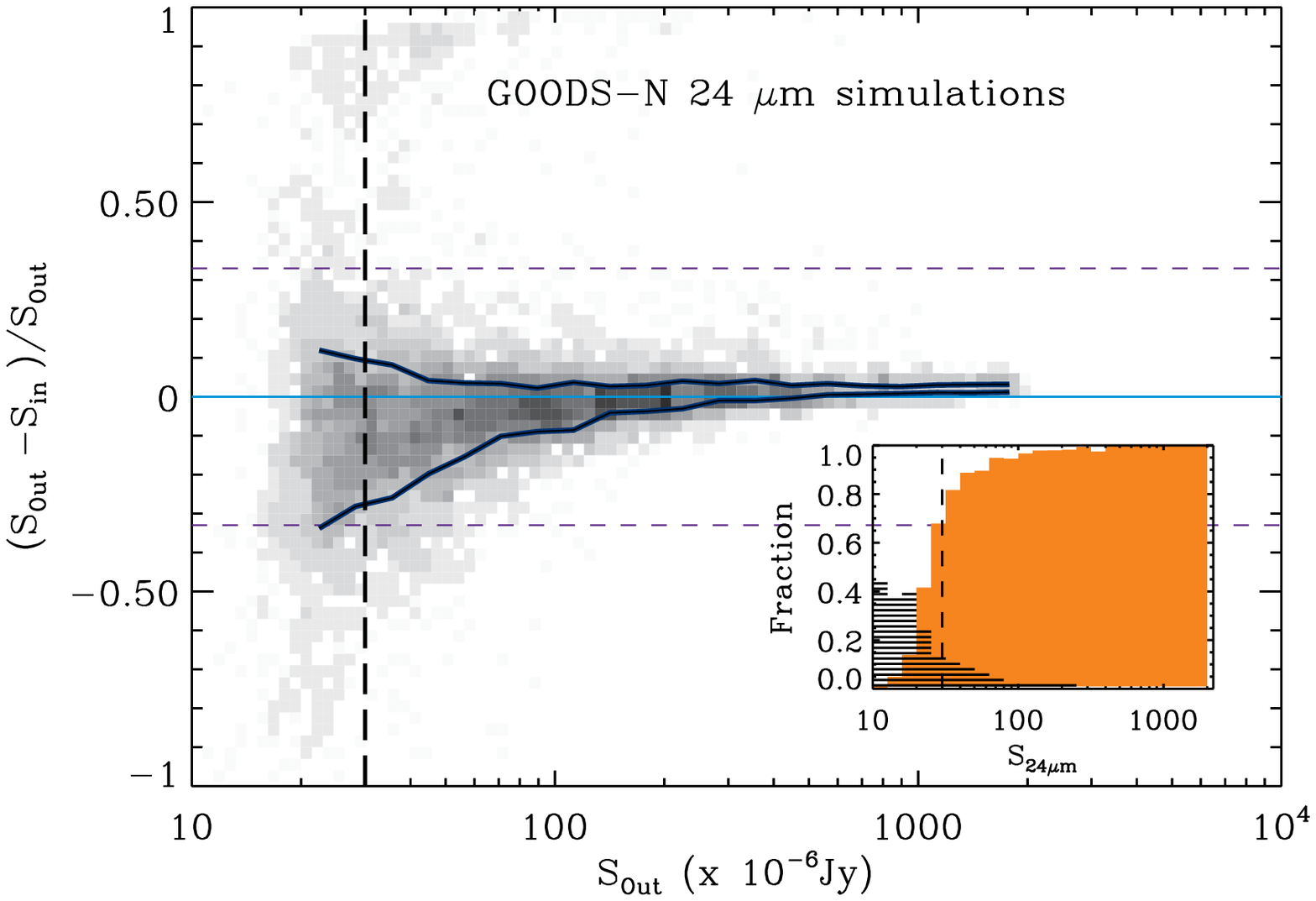}
	\includegraphics[width=9.cm]{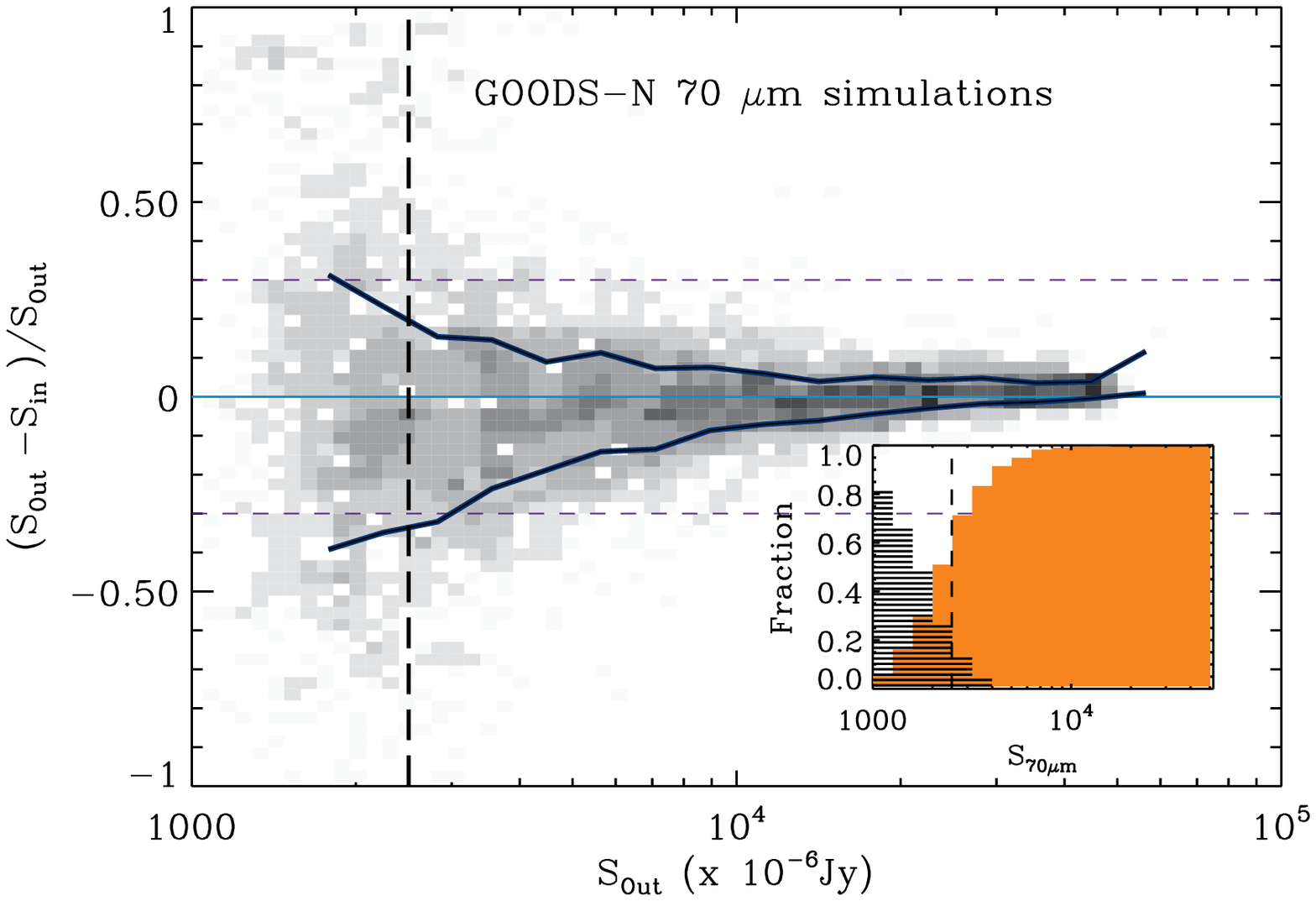}
	\caption{\label{fig:simulation}Photometric accuracy, [S(out)-S(in)]/S(out), derived from the simulation as a function of measured flux density, S(out), for the GOODS-N 24 $\mu$m (\textit{left}) and 70 $\mu$m observations (\textit{right}).
Shaded gray region shows the distribution that we obtain and the solid black lines contain 68 \% of the galaxies.
Vertical long dashes lines show the limit of our catalog defined as the best balance between photometric accuracy, completeness and contamination.
Inset plots show the fraction of artificial sources detected in the image (completeness) as a function of input flux (\textit{orange plain histogram}) and the fraction of \textit{spurious} sources (contamination) as a function of flux density (\textit{striped black histogram}).
For the definition of the completeness and contamination see the text.
}
\end{figure*}
\subsubsection{GOODS-North $\&$ South\label{subsubsec:GOODS-N/S}}
\indent{
Infrared imaging of the GOODS-N field ($12^{h}36^{m},\,+62^{\circ}14\arcmin$) was obtained at 24 $\mu$m as part of the GOODS Legacy program (PI: Dickinson).
The 70 $\mu$m data were obtained as part of the GO-3325 \citep[PI:][]{frayer_2006} program for the $10\arcmin\times10\arcmin$ central region and as part of the FIDEL legacy program for the northeast and southwest regions in order to cover the entire GOODS-N area (i.e $15\arcmin\times10\arcmin$).
Infrared imaging of the GOODS-S field ($3^{h}32^{m},\,-27^{\circ}48\arcmin$) was obtained at 24 and 70 $\mu$m as part of the GOODS legacy program (PI: Dickinson) and the GO-20147 (PI: Frayer) program respectively.
We note that even if the GOODS-S 24 $\mu$m field covers a $10\arcmin\times18\arcmin$ area, the deep 70 $\mu$m observations only cover a $10\arcmin\times10\arcmin$ area.
Thus the GOODS-N and the GOODS-S fields cover respectively a total area of 194 arcmin$^{2}$ and 91 arcmin$^{2}$.
\\}
\indent{
Final mosaics, obtained using MOPEX, have the same pixel scale as the EGS (i.e 1.2$\arcsec/$pixel at 24 $\mu$m and 4.0$\arcsec/$pixel at 70$\mu$m).
In both fields observations reach point source sensitivity of $\thicksim30\ \mu$Jy and $\thicksim2\,500\ \mu Jy$ at 24 and 70 $\mu$m respectively (i.e an effective integration time per sky pixel of $\thicksim35\,000$s and $\thicksim10\,800$s respectively).
Source extraction was carried out using our method with prior positions, leading to a total of 2151 sources at 24 $\mu$m($\ge 30\,\mu$Jy, 5$\sigma$) and 119 sources at 70 $\mu$m ($\ge 2.5\,$mJy) in GOODS-N and 870 sources at 24 $\mu$m ($\ge 30\,\mu$Jy) and 44 sources at 70 $\mu$m ($\ge 2.5\,$mJy) in GOODS-S. 
\\}
\subsubsection{ECDF-S}
\indent{
The ECDF-S ($3^{h}32^{m},\,-27^{\circ}48\arcmin$) was observed with the MIPS instrument as part of the FIDEL legacy program (PI: Dickinson) and covers a total area of 900 arcmin$^{2}$.
In the following we refer to the ECDF-S field as the outer part of this $30\arcmin\times30\arcmin$ area while the central part ($10\arcmin\times10\arcmin$) is refered to as the GOODS-S field (see Section \ref{subsubsec:GOODS-N/S}).
Reduced to its outer part, the ECDF-S field covers a total area of 520 arcmin$^{2}$ (hereafter ECDFS-O).
The choice to treat separately the GOODS-S and the ECDFS-O data is driven by the difference in the 70 $\mu$m coverage (the central part being deeper than the outer part).
The final 70 $\mu$m mosaic, obtained using MOPEX, has the same pixel scale as the EGS and reaches a point source sensitivity of $\thicksim3.5$ mJy (i.e an effective integration time per sky pixel of $\thicksim5\,750$s).
At 24 $\mu$m, the effective integration time per sky pixel being inhomogeneous, the final mosaic reaches point source sensitivity values which are function of the source positions.
It spans the range of $\thicksim40\,-\,70\ \mu$Jy over the whole mosaic.
\\}
\indent{
Using our extraction method we obtained 2185 sources at 24 $\mu$m ($\ge 80\,\mu$Jy) and 195 sources at 70 $\mu$m ($\ge 3.5\,$mJy).
}
\subsection{Infrared source detection \label{subsec: prior method}}
\indent{
Since most sources in these fields were unresolved, we performed a PSF fitting technique to extract their photometry.
Nevertheless, to be able to deal with blending issues and to go down to the limit of the data, we used, as a prior information, the expected position of the sources.
We assume that all sources present in the mid-infrared images have already been detected at other wavelengths where the depth and the resolution of the data are higher.  \\
}
\indent{
For the 24 $\mu$m data, we choose to use the position of the IRAC-3.6 $\mu$m sources.
This choice was motivated by the depth of the IRAC-3.6 $\mu$m data which is on average a factor of 30 deeper than MIPS 24 $\mu$m data in every field used in this study.
Hence since we know that the typical $S_{24\,\mu m}/S_{IRAC}$ ratio spans the range [3-20] \citep{chary_2004}, the IRAC data are deep enough to contain every 24 $\mu$m source.
Moreover the resolution of the IRAC-3.6 $\mu$m data (FWHM $\thicksim 2 \arcsec$) being 3 times sharper than the MIPS 24 $\mu$m data, blended sources separable in the 24 $\mu$m MIPS images (i.e sources separated by more than $\thicksim 2 \arcsec$ in the 24 $\mu$m data) are unblended in the IRAC catalog.
\\}
\indent{
For the GOODS-N field, we used the publicly available IRAC catalog released as part of the GOODS legacy program (Dickinson et al., in preparation).
This catalog contains $19\,437$ objects detected at 3.6 $\mu$m with a 50\% completeness limit of 0.5 $\mu$Jy and an astrometric accuracy of $0.37\arcsec$. 
For the EGS, we used the publicly available IRAC EGS catalog released by the AEGIS collaboration \citep{barmby_2008}.
This catalog contains 57,434 objects detected at 3.6 $\mu$m with a 50\% completeness limit of 1.5 $\mu$Jy.
The astrometric accuracy of this catalog is estimated to be close to $0.37\arcsec$. 
Finally, for the GOODS-S/ECDF-S region, we used the SIMPLE  catalog \citep[\textit{Spitzer} IRAC / MUSYC Public Legacy in ECDF-S;][]{gawiser_2006}.
This catalog contains $61\,233$ objects detected at 3.6 $\mu$m with a 50\% completeness limit of 1.5 $\mu$Jy.
The astrometric accuracy is the same as in the EGS catalog.
\\}
\begin{table*}
\label{tab:catalog} 
\center{
\caption{Catalog properties}
\renewcommand{\footnoterule}{} 
\begin{tabular}{ccccccccc}
\hline \hline
Field & Area & Exposure & Flux Limit$^{\,\mathrm{a}}$ & Completeness$^{\,\mathrm{b}}$ & Contamination$^{\,\mathrm{b}}$ & \# sources \\
 & \tiny {($\rm{arcmin^{2}}$)} & \tiny {($\rm{sec/pixel}$)} & \tiny {($\rm{\mu Jy}$)}& \tiny{($\rm{\%\ at\ S=S_{limit}}$)} &  \tiny{($\rm{\%\ at\ S=S_{limit}}$)}  &    \tiny{($\rm{with\ S \ge S_{limit}}$)} \\
 \hline
EGS 24 & 548 &14\,000 & 50 (5$\sigma$) & 75 & 11 & 4096\\
GOODS-S 24 & 91 & 35\,000 & 30 (5$\sigma$)& 81 & 13 & 870 \\
GOODS-N 24 & 194 & 35\,000 & 30 (5$\sigma$)& 81 &  13 & 2151\\
ECDFS-O 24 & 517 & 8\,000 & 70 (5$\sigma$)& 100 &  0 & 2474 \\
\\
EGS 70 & 548 & 7\,200 & 3000 (6$\sigma$)& 77 & 10 & 322\\
GOODS-S 70 & 91 & 10\,800 & 2500 (6$\sigma$)& 73 & 12 & 44 \\
GOODS-N 70 & 194 & 10\,800 & 2500 (6$\sigma$)& 73 &  12 & 119 \\
ECDFS-O 70 & 517 & 5\,750 & 3500 (6$\sigma$)& 85 &  7 & 195 \\
\hline
\end{tabular}
}
\begin{list}{}{}
\item[$^{\mathrm{a}}$]  the level of the noise was calculated on the residual images and hence it combines the confusion, instrumental and photon noises.
\item[$^{\mathrm{b}}$]  see text for the definition of the completeness and contamination
\end{list}
\end{table*}
\indent{
An empirical 24 $\mu$m PSF was constructed with isolated point like objects present in the mosaic and was then fitted to all IRAC positions in the map.
The photometry of each 24 $\mu$m source is defined as the scaled fitted PSF.
Then, as for a standard aperture correction, the photometry is corrected to account for the finite size of the empirical PSF.
A residual image was created by subtracting from the original image the detected 24 $\mu$m sources.
Finally we check in this residual image the presence of 24 $\mu$m sources missed by the lack of IRAC position using DAOPHOT detection.
We find no residual sources.
\\}
\indent{
For the 70 $\mu$m data, we choose to start from our previous 24 $\mu$m detections.
This choice is straightforward since the typical $S_{70\,\mu m}/S_{24\,\mu m}$ ratio spans the range [2-100] \citep{papovich_2007} and that the 24 $\mu$m observations are about 100 times deeper.
No reliable empirical PSF can be constructed at 70 $\mu$m since only a few isolated sources can be found in each map.
We decided to use the appropriate 70 $\mu$m Point Response Function (PRF) estimated on the extragalactic First Look Survey \citep[xFLS;][]{frayer_2006_b} mosaic and available on the \textit{Spitzer} web site.
Then as for the 24 $\mu$m detection, we fit the 70 $\mu$m maps with this synthetic PSF at each 24 $\mu$m position.
Once again we then check the residual image but no residual sources where found, i.e we did not see any evidence for the existence of 70 $\mu$m sources with no 24 $\mu$m counterpart at these depths.
\\}
\indent{
The advantage of this method of priors is principally to resolve a large part of the blending issue.
Moreover this method reduces the cross-matching issue since for each IRAC position we directly have a 24 and 70 $\mu$m flux.
Then to cross-match our mid-infared detections with any other catalog, we can use the IRAC astrometry which is more accurate than any 24 or 70 $\mu$m centroid position.
\\ \\}
\indent{
To estimate completeness and photometric reliability we performed extensive simulations for each field and at each wavelength.
We added artificial sources in the image with a flux distribution matching approximately the measured number counts \citep[see][]{frayer_2006, papovich_2004}.
In order to preserve the original statistics of the image (especially the crowding properties) the numbers of artificial objects added in the image was kept small (as an example for GOODS-N we only added 40  sources into the 24 $\mu$m image and 4 sources into 70 $\mu$m images). 
Moreover as the photon noise of the MIPS 24 and 70 $\mu$m data is dominated by the background emission we do not need to introduce any additional Poisson noise due to the photon statistics of the object itself.
 We then performed the source extraction again to compare the resulting photometry to the input values.
To increase the statistics we used repeatedly the same procedure with different positions in the same field.
For each field we introduced a total of $20\,000$ artificial objects.
\\}
\indent{
In Figure \ref{fig:simulation} we show the results of one of the simulations obtained for the GOODS-N 24 $\mu$m and 70 $\mu$m data.
We plot for each wavelength the photometric accuracy derived from the simulation as a function of measured flux density.
We defined the completeness as the fraction of sources with a flux accuracy better than 50\%.
To estimate the contamination of our catalog by spurious sources (i.e sources arising from the noise properties of the image and not from real sources) we simulated very faint sources which were normally undetectable since they were introduced with a flux density fainter than the background noise.
The contamination was then defined as the fraction of these undetectable sources that were extracted by our method.
Using these simulations we defined the flux density limits of the final catalogs as those above which we simultaneously get :
\\}
\indent{
- a photometric error better than 33\% for at least 68\% of the sources.
\\}
\indent{
- a completeness of at least $\thicksim80\%$ (to minimize future completeness correction).
\\}
\indent{
- a contamination by artificial sources lower than 15 \%.
\\}
\indent{
Under this definition, we found that the GOODS-N and S 24 $\mu$m catalogs, which have about the same exposure time, reach a limit of 30 $\mu$Jy (see Table \ref{tab:catalog}).
We emphasize that sources as faint as $\thicksim 15-20\,\mu$Jy can be extracted from the 24 $\mu$m images.
However, in order to avoid contamination and large incompleteness, both of which are significant systematics in the estimation of the LF, we have cut the GOODS 24 $\mu$m catalogs at a brighter flux density limit.
For the EGS 24 $\mu$m data the flux density limit of the catalog is 50 $\mu$Jy.
The 24 $\mu$m exposure map of the outer part of the ECDF-S (i.e ECDFS-O)  is inhomogeneous, hence we used the lowest exposure time as the reference to define the flux density limit.
In order to avoid inhomogenous completeness corrections we imposed a level for the flux density limit such that the completeness is $\thicksim100\%$.
Combining these two conditions, the ECDFS-O 24 $\mu$m catalog reaches a limit of 70 $\mu$Jy.
\\}
\indent{
For the 70 $\mu$m data of GOODS-N and S, the catalog flux density limit is $2.5\,$mJy.
For the EGS 70 $\mu$m data, the flux limit of our catalog is $3\,$mJy and for the ECDFS-O 70 $\mu$m data the detection limit of the catalog is $3.5\,$mJy.
\\}
\indent{
We would like to emphasize that the calibration factor taken to generate the final 24 and 70 $\mu$m mosaics are derived from stars, whose SED at these wavelengths are generally very different from those of distant galaxies.  
Hence color corrections may be necessary to these fluxes (at most $\thicksim 10\%$).
Nevertheless, since these color corrections are highly dependent on the redshift and the intrinsic SED of the sources, we decided to introduce these color corrections directly in the \textit{k}-correction used to estimate the LF (see Section \ref{subsec:methodo 15}) since both quantities are taken into account in this computation.
\\}
\subsection{Spectroscopic redshifts}
\subsubsection{EGS}
\indent{
The optical spectroscopic redshifts of the EGS field were taken from the DEEP2 data release 3 catalog \citep{davis_2007}.
This survey targeted $\sim$50,000 distant galaxies with $R<24.5$, using the DEIMOS spectrograph on the Keck II telescope.
In the EGS, unlike the other DEEP2 fields, objects at $z < 0.7$ were not excluded from the selection.
The data release 3 catalog contains about $31\,600$ reliable redshifts with $18.5<R<24.5$.
For more details on the selection criteria of the DEEP 2 program see \citet{davis_2007}.
\\}
\begin{table*}
\caption{\label{tab:catalog final} Redshift catalog properties}
\center{
\begin{tabular}{ccccccccc}
\hline \hline
Field & Area & Nb  sources & X-ray AGN & No X-ray AGN & \# spec-$z$$^{\,\mathrm{a}}$ & \# phot-$z$$^{\,\mathrm{a,b}}$ & \# spec-$z$ and/or phot-$z$$^{\,\mathrm{a}}$ \\
 & \tiny {($\rm{arcmin^{2}}$)}  & \tiny {$\rm{24 \mu m/ 70 \mu m}$} &  \tiny{$\rm{24 \mu m/70 \mu m}$}  &  \tiny{$\rm{24 \mu m/70\mu m}$}  & \tiny{$\rm{24 \mu m/70 \mu m}$}  & \tiny{$\rm{24 \mu m/70 \mu m}$}  & \tiny{$\rm{24 \mu m/70 \mu m}$} \\
\hline
EGS & 548 & $4096/322$ & $181/27$ & $3915/285$ & $1108/128$ & $1773/126$ & $2881/254$\\
& &  & \tiny{$4\%/11\%$} & \tiny{$96\%/89\%$} & \tiny{$28\%/45\%$} & \tiny{$45\%/45\%$} & \tiny{$73\%/90\%$}\\
\\
GOODS-S & 91 & $870/44$ & $64/7$ & $806/37$ & $371/32$ & $378/5$ & $749/37$\\
 & &  & \tiny{$7\%/16\%$} & \tiny{$93\%/84\%$} & \tiny{$46\%/86\%$} & \tiny{$46\%/14\%$} & \tiny{$92\%/100\%$}\\
\\
GOODS-N & 194 & $2151/119$ & $134/12$ & $2017/107$ & $747/72$ & $1148/29$ & $1895/101$\\
&  &  & \tiny{$6\%/10\%$} & \tiny{$94\%/90\%$} & \tiny{$37\%/67\%$} & \tiny{$57\%/27\%$} & \tiny{$94\%/94\%$}\\
\\
ECDFS-O & 517 & $2474/195$ & $0/0$ & $2474/195$ & $180/16$ & $1288/168$ & $1468/184$\\
&  &  & \tiny{$0\%/0\%$} & \tiny{$100\%/100\%$} & \tiny{$7\%/8\%$} & \tiny{$52\%/86\%$} & \tiny{$59\%/94\%$}\\
\hline
\end{tabular}
}
\begin{list}{}{}
\item[$^{\mathrm{a}}$] the percentage note in these columns refer to the number of no X-ray AGN
\item[$^{\mathrm{b}}$] number of sources which have a photometric redshift but no spectroscopic redshift
\end{list}
\end{table*}
\subsubsection{GOODS-North/South $\&$ ECDF-S}
\indent{
A large number of spectroscopic redshifts has been measured for galaxies in the GOODS regions.
In GOODS-N a total of 2376 spectroscopic redshifts come from a combination of various studies \citep[][Stern et al. in prep]{cohen_2000,wirth_2004,cowie_2004}.
In the GOODS-S/ECDF-S region, 2547 spectroscopic redshifts were also combined from various studies \citep[][]{vanzella_2006,lefevre_2004,mignoli_2005}.
Although these spectroscopic redshift catalogs were obtained from several campaigns with diverse selection criteria, when combined together, they provide a rather homogeneous sampling.
In GOODS-N and GOODS-S, respectively 60\% and 50\% of the galaxies brighter than $z_{AB}=23.5$ and with a photometric redshift lower than $z_{phot}<1.2$, have a spectroscopic redshift \citep[see][]{elbaz_2007}.
\\}
\subsection{Photometric redshifts}
\label{subsec:photometric redshift}
\subsubsection{EGS}
\indent{
In addition to the spectroscopic redshifts we also need reliable photometric redshifts.
We used publicly available photometric redshifts provided by the TERAPIX and the VVDS consortia \citep{ilbert_2006} using the Canada-Hawai-France Telescope Legacy Survey (CFHT-LS).
The CFHTLS D3 survey consists of a 1 square degree region centered at $\alpha=14^{h}19^{m}27^{s},\delta=+52^{\circ}40\arcmin65\arcsec$ observed through the \textit{u, g, r, i} and \textit{z} CFHTLS filters.
Using this set of photometric magnitudes and appropriate spectral energy distributions (SED), the "Le Phare" photometric code provided a redshift probability distribution for each source \citep{ilbert_2006}.
\\}
\indent{
The final $i\arcmin_{AB}$ selected catalog contains $366\,030$ sources with a magnitude limit of $i\arcmin_{AB} \thicksim26.0$.
For galaxies with $i\arcmin_{AB}<24$ and $z<1.5$, these redshifts have an accuracy of $\sigma_{\Delta z/(1+z)}=0.029$ with 3.8\% of catastrophic errors \citep{ilbert_2006}.
As recommended in \citet{ilbert_2006}, we excluded all sources with a double peak in their redshift probability distribution function (PDF) since most of them did not have a reliable redshift. 
The overlap region between the EGS and the CFHTLS D3 reduces the EGS area from 900 arcmin$^{2}$ down to 548 arcmin$^{2}$.
\\}
\subsubsection{GOODS-North $\&$ South}
\indent{
For the GOODS fields, we complemented the spectroscopic redshifts with photometric redshifts estimated using Z-PEG \citep{leborgne_2002}.
For both fields we used some of the deepest data currently available in the GOODS region (\textit{UBVRIzJHK}, 3.6 $\mu$m,  4.5 $\mu$m) to derive photometric redshifts with an accuracy of $\sigma_{\Delta z/(1+z)}=0.05-0.1$.
For more detail on the complete multi-lambda catalog used in the computation of the photometric redshifts see Le Borgne et al. (In Prep).
\\}
\subsubsection{ECDF-S}
\indent{
For the ECDF-S we used the catalog of photometric redshifts from the COMBO-17 survey \citep[Classifying Objects by Medium-Band Observations in 17 filters ;][]{wolf_2004}.
The COMBO-17 survey observed a $31.5\,\times\,30$ arcmin$^{2}$ area with 5 broadband and 12 narrowband filters.
The catalog contains $63\,501$ objects selected on the deep R-band image ($R<26$) carried out with the Wide Field Imager (WFI) at the MPG/ESO 2.2-m telescope in La Silla.
Using an extensive SED library, they fitted the photometry of each object to compute its redshift probability distribution.
Finally the COMBO-17 catalog contains 24\,216 objects with a reliable photometric redshift.
\\}
\indent{
For sources with $R<24$ and $z<1.2$, these redshifts have an accuracy of $\sigma_{\Delta z/(1+z)}=0.1$.
For sources fainter than $\rm{R}=24$ the redshift accuracy drops dramatically, therefore we did not consider in the following such faint objects.
\\}
\subsection{Removing Active galaxies}
\indent{
The infrared luminosity of a normal star-forming galaxy can be severely affected by the presence of an Active Galactic Nucleus (AGN).
Hence to study the star formation history of normal galaxies it is necessary to identify and remove these AGNs.
\\}
\indent{
In order to test the presence of AGNs in the GOODS field, we used the deepest \textit{Chandra} X-ray observations i.e the 1Ms maps for the CDF-S and the 2 Ms maps  for the CDF-N \citep{alexander_2003}.
For both fields we used the corresponding public X-ray catalog.
For the EGS field we also used \textit{Chandra} observations taken from the public AEGIS-X data release 2 point source catalog which has a 200 Ksec coverage \citep{laird_2008}.
Finally for the ECDF-S we used the 250 Ksec \textit{Chandra} observations, which flank the 1 Ms CDF-S observations \citep{lehmer_2005}.  
\\}
\indent{
 AGNs were identified as galaxies with either $L_{X}\, [0.5\,-\,8.0\ keV]\,\ge\,3\,\times\,10^{42}\ \rm{erg\, s^{-1}}$ and/or a hardness ratio greater than 0.8 (ratio of the counts in the $2 - 8$ keV to $0.5-2$ keV passbands) \citep{bauer_2004}.
 As mentioned in \citet{bauer_2004}, some AGNs might be missed by this technique.
The fraction of such galaxies remains a matter of debate until today but the global consistency of the 24-70 $\mu$m color of the galaxies studied here (as discussed in Section \ref{subsec:stacking}) suggests that they should not play a dominant role \citep[see also][]{fadda_2002}.
  We did not subtract the AGN contribution to the IR light of those galaxies harboring an AGN, instead we conservatively assumed that any galaxy contaminated by an AGN was not forming stars. It is well-known that AGNs do also harbor star formation, especially those emitting strongly at IR wavelengths, but such subtraction would be highly speculative at the present level of our knowledge.
\\}
\subsection{The final IR galaxy sample}
\indent{
To construct our final catalog, we first truncated our mid-infrared catalog to the region covered by both \textit {Spitzer} and the optical redshift surveys used in this study.
This step was especially important for the EGS field where the total FIDEL area of 900 arcmin$^{2}$ is reduced to 548 arcmin$^{2}$ to overlap the CFHT-LS survey (i.e the photometric redshift survey).
We then cross-matched the 24 $\mu$m catalog with the X-ray observations to exclude AGNs.
We found that $\thicksim6 \%$ ($\thicksim10 \%$) of the 24 $\mu$m (70 $\mu$m) sources were affected by an X-ray AGN and hence removed from our catalog.
The rest of the data was then cross-matched with the different spectroscopic catalogs and finally the remaining sources were matched with the photometric catalogs.
All these cross-matching steps were made using the IRAC astrometry of the mid-infrared sample and a matching radius of 1.5 arcsec.
The choice of 1.5 arcsec is motivated by the size of the IRAC-$3.6\,\mu$m PSF FWHM ($\sim 2\arcsec$).
In case of multiple associations we selected the closest optical counterparts to the IRAC centroid.
We found that 8\%, 10\% and 9\% of the MIPS 24 $\mu$m sources have multiple associations with optical sources brighter than $\rm{I}<26$ mag for the EGS, GOODS-N, GOODS-S.
For the ECDFS-O we found that 1\% of sources have a multiple association with optical sources brigther than $\rm{R}<24\,\rm{mag}$.
\\}
\indent{
All the different steps described previously and applied to the different fields of our study are listed in the Table \ref{tab:catalog final}. The redshift completeness is larger than 90\,\% (70\,\% except for the ECDFS-O) at 70\,$\mu$m (24\,$\mu$m) for the sources with no AGN contribution.
\\}
\subsection{Redshift uncertainties and sample completeness\label{subsec: uncertainties}}
\indent{
One of the major concern of all studies heavily relying on photometric redshift is their possible redshift incompleteness for distant and/or faint galaxies.
As these issues may have a nonnegligible effect when estimating luminosity functions as a function of redshift, one should pay attention to the consequence of these limitations.
\\ }
\begin{figure}
	\includegraphics[width=8.5cm]{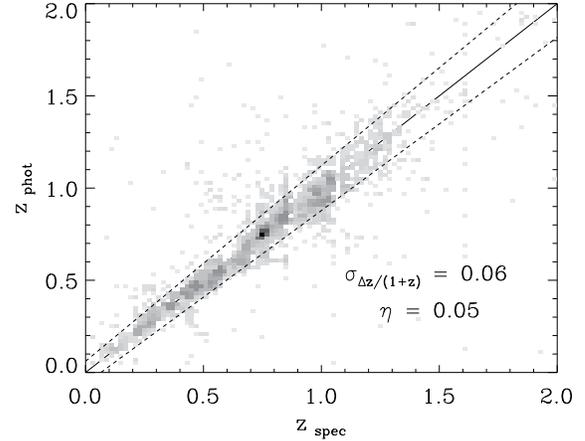}
	\caption{\label{fig: spectro photo}Comparison between photometric and spectroscopic redshift for 2406 galaxies of our 24 $\mu$m-selected catalog.
	Dashed lines represent the relative errors (i.e $\sigma_{\Delta z/(1+z)}=0.06$).
	$\eta$ gives the fraction of source with $\Delta z/(1+z)>0.15$. 
	}
\end{figure}
\begin{figure}
	\includegraphics[width=8.cm]{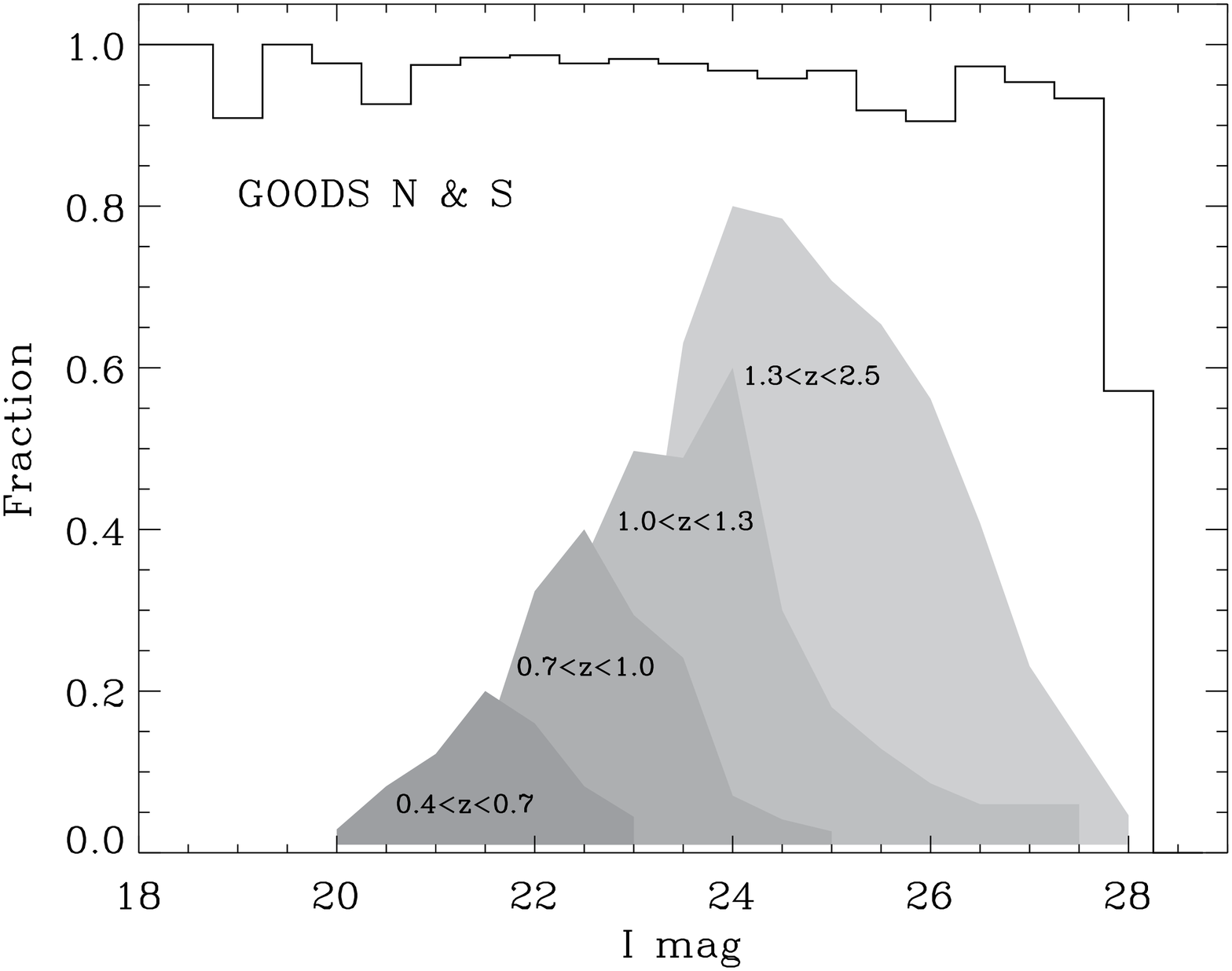}
	\includegraphics[width=8.cm]{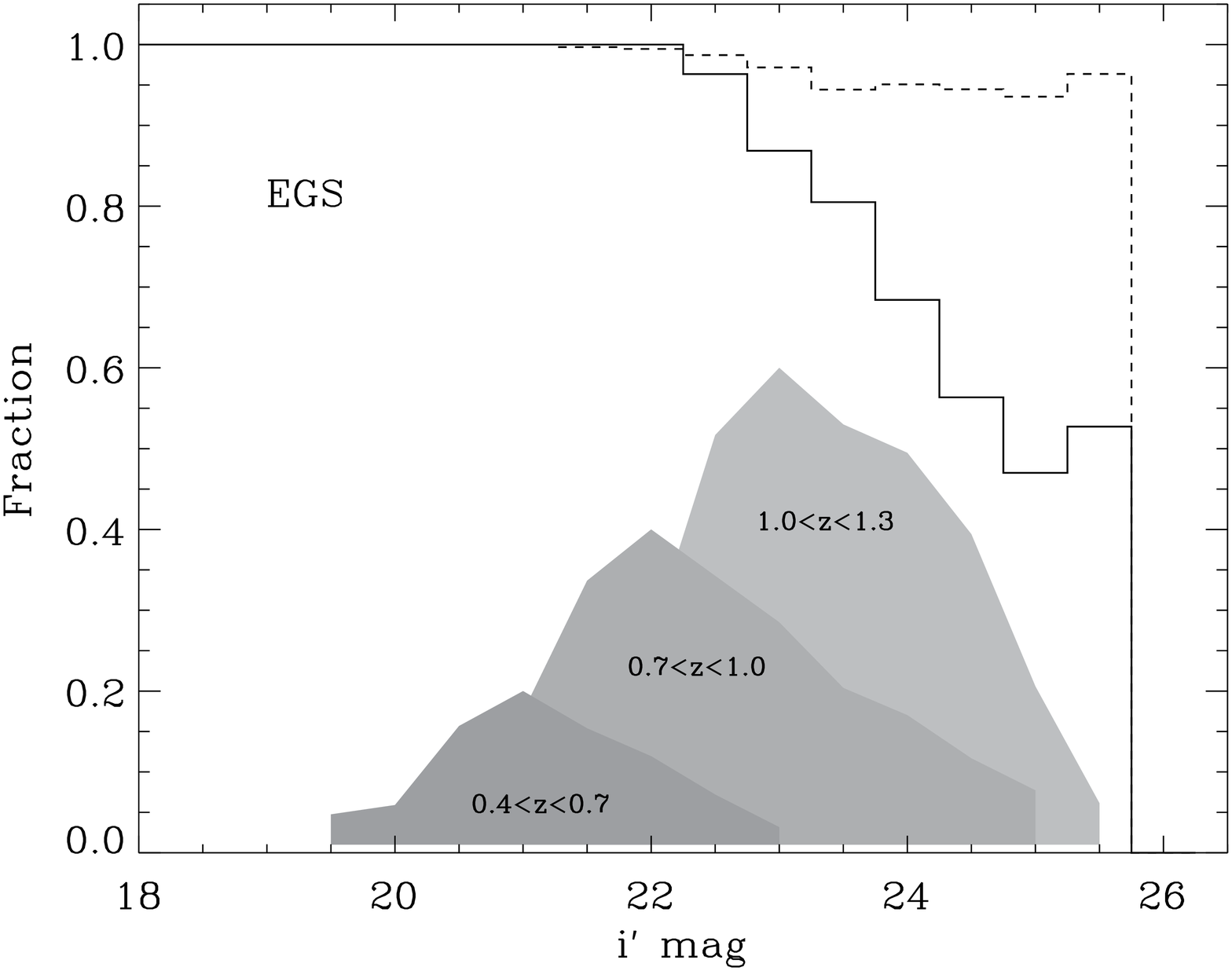}
	\includegraphics[width=8.cm]{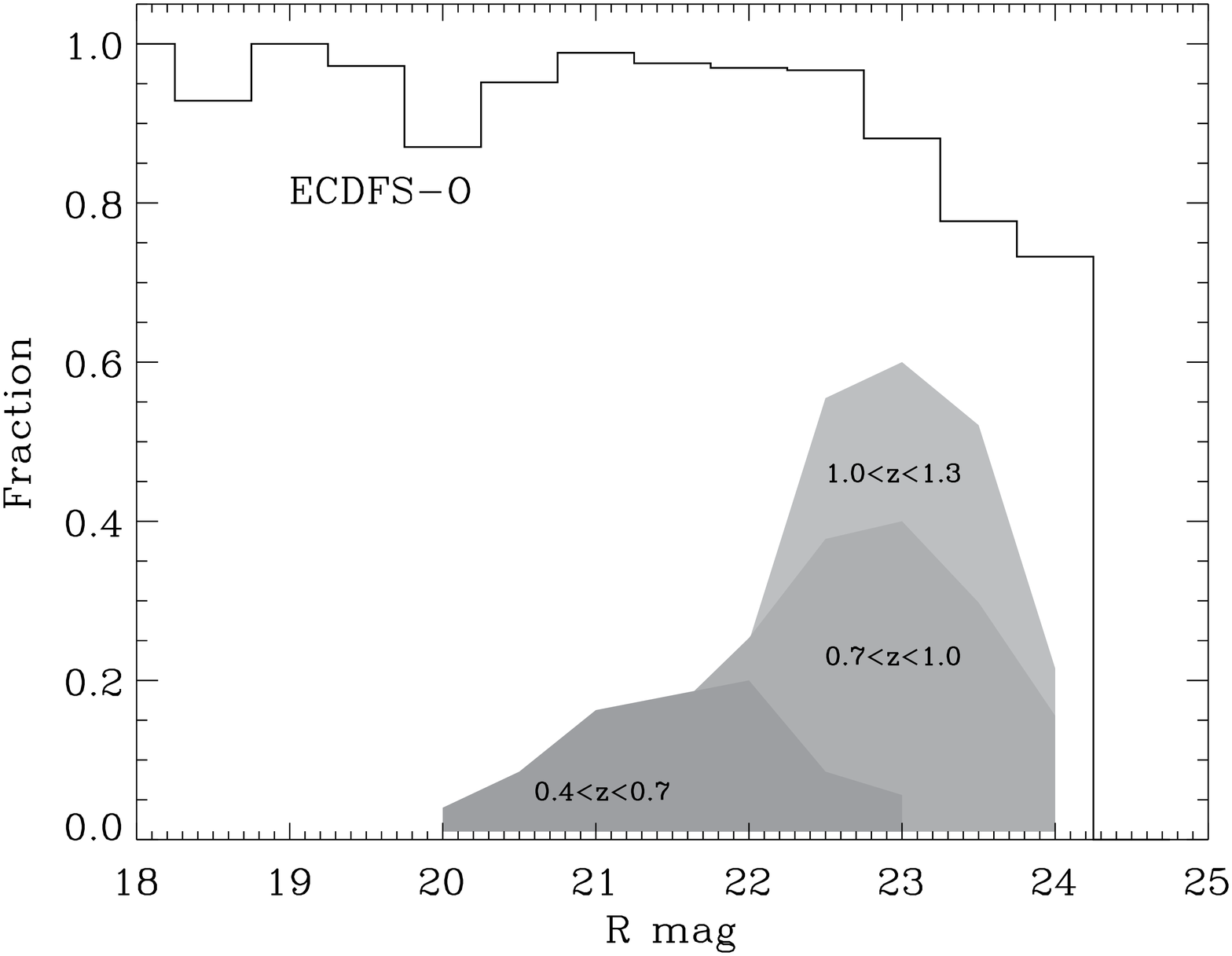}
	\caption{\label{fig: optical} Fraction of 24 $\mu$m sources with an optical counterpart which have a reliable photometric redshift as function of optical magnitude (\textit{solid line}).
	The \textit{dashed line}, in the EGS panel, represents the redshift determination completeness as a function of the optical magnitude before the exclusion of all sources with a double peak in their PDF (see Section \ref{subsec:photometric redshift}).
	The shaded regions show the optical magnitude distribution for various redshift bins (all histograms have been scaled with an arbitrary factor). Note that the ECDFS-O panel uses the R-band magnitude.	
	}
\end{figure}
\indent{
In Section \ref{subsec:photometric redshift} we have quoted the accuracy of all photometric redshift catalogs used in this work.
Nevertheless since these accuracies were based on an optically-selected sample we need to characterize the typical uncertainties of our 24 $\mu$m-selected sample.
In Figure \ref{fig: spectro photo} we compare the spectroscopic and photometric redshifts of all 24 $\mu$m sources with a spectroscopic redshift after combining the four fields (2406 sources, see Table~ \ref{tab:catalog final}).
We observe a very good agreement between the photometric and spectroscopic redshifts with an accuracy of $\sigma_{\Delta z/(1+z)}=0.06$ and a median value of $-0.005$.
We notice that this accuracy is not statistically larger than the accuracy quoted for the optically-selected sample.
We also found $4\%$ of catastrophic objects defined as those objects with $\Delta z/(1+z)>0.15$.
Even if those galaxies are statistically rare, they could have a major effect on the estimates of the LF by boosting the bright end slope.
Therefore in the following study we will pay close attention to quantify their effect on the estimates of the total infrared LF (see Section \ref{subsec:methodo 15}).
\\}
\begin{figure}
	\includegraphics[width=8.5cm]{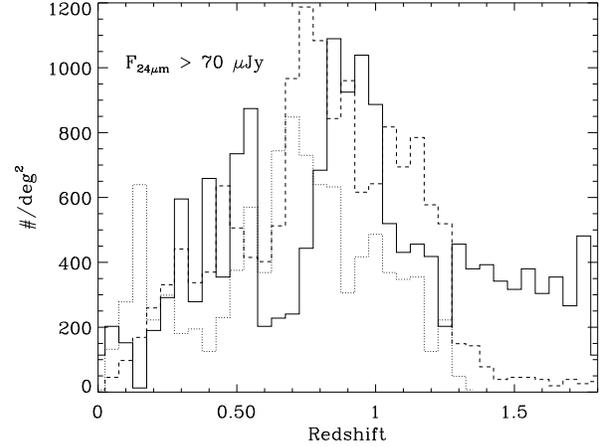}
	\caption{\label{fig: distri} Redshift distributions of 24 $\mu$m sources selected with $F_{24\,\mu\rm{m}}>70\,\mu\rm{Jy}$  in the GOODS-S and N fields (solid line), in the EGS field (dash line) and in the ECDFS-O field (dot line).
	}
\end{figure}
\indent{
As previously mentioned and quoted in Table \ref{tab:catalog final} a high fraction ($\thicksim 80\%$) of our final IR galaxy sample have been associated with a spectroscopic and/or a photometric redshift.
Nevertheless the remaining fraction of sources with no redshift association are not randomly distributed in the redshift space.
As a result, we need to quantify the redshift above which the photometric redshift incompleteness affects our study.
In Figure~\ref{fig: optical} we present the photometric redshift completeness as a function of the optical magnitude (\textit{solid lines}).
These histograms allow us to estimate the optical magnitude above which the determination of a reliable photometric redshift start to be problematic.
Independently, we also present in figures \ref{fig: optical} the optical magnitude distribution of the 24 $\mu$m sources for different redshift bins.
For clarity these optical magnitude distributions have been scaled with an arbitrary factor.
By combining these two informations (i.e the redshift determination completeness as a function of the optical magnitude and optical magnitude distribution) one can have an estimates of the redshift incompleteness of his sample.
For example, if the photometric redshift catalog is highly incomplete for sources fainter than $R>24$ and in the same time a large fraction of the 24 $\mu$m sources that lie at $1<z<1.3$ have $R\thicksim24.5$, one can expect a large redshift incompleteness in this redshift bin.
\\}
\indent{
As already discussed in \citet{lefloch_2005} we observe that sources situated at higher redshift have fainter optical counterpart and that the redshift determination completeness decreases at faint magnitudes.
For the GOODS fields we notice that the depth of optical observations and the use of deep near infrared data allow us to have a complete redshift association up to $z\thicksim2$.
Indeed we notice that the redshift completeness of the GOODS field falls at optical magnitudes (i.e $R>27.5$) where the distribution of 24 $\mu$m sources that lie at $1.3<z<2.5$ is already low.
Therefore for the GOODS fields, sources with no redshift association ($\thicksim5\%$) are likely located at $z>2$ and hence should not affect the study of the evolution of the total infrared LF up to $z=1.3$. 
For the EGS the redshift determination completeness highly decreases at an optical magnitude range (i.e I-mag$\thicksim22.5 $) where a large fraction of sources are located at $z<1.3$.
This trend can lead to a high incompleteness of sources located at $1<z<1.3$.
However we notice that the decrease of this redshift determination is driven by sources with a double peak in their probability distribution functions (PDF; see the \textit{dash lines} in Figure \ref{fig: optical}). All galaxies for which we have both a double peak distribution in their photometric redshift and a spectroscopic redshift are located at $z>1$.
The minor effect of "double peak" galaxies on the redshift distribution can be inferred by the comparison of the redshift distribution of the EGS and the GOODS fields (for which we expect no redshift incompleteness up to $z\thicksim 2$; see Figure \ref{fig: distri} and Figure~\ref{fig: optical}).
Indeed we do not observe large differences in the redshift distribution of these two fields between $0<z<1.3$, whereas at $z>1.3$ the number of sources in the EGS drops to zero likely due to redshift incompleteness.
Finally for the ECDFS-O field we find the same results than \citet{lefloch_2005}, i.e the magnitude distribution of the two highest redshift bins (i.e $0.7<z<1.0$ and $1.0<z<1.3$) are truncated at the faint end which indicate that these redshift bins are affected by incompleteness.
Nevertheless we can estimate from the shape of the magnitude distribution that only $\thicksim15\%$ of sources should be affected by this incompleteness and hence no large effects are expected on the estimates of the infrared LFs.
Indeed we will see in the following that the infrared LFs inferred from the ECDFS-O and GOODS fields are, within the error bars, in very good agreement (see Section \ref{sec:15 LF}, \ref{section: function 35} and \ref{sec:ir LF}).
To conclude we infer that the redshift incompleteness should not affect the conclusions of our study since the uncertainties corresponding to this misidentification are lower than the uncertainties introduced by the \textit{k}-correction and the photometric redshift measurements.
\\}
\indent{
Assuming that our photometric redshift catalogs do not suffer from incompleteness up to $z=1.3$, we can construct a reference sample of galaxies located at $z<1.3$.
When compared to this reference sample, we find that the fraction of sources with a spectroscopic redshift is $56\%$, $35\%$ and, $9\%$ for the GOODS, the EGS and the ECDFS-O fields respectively.
\\}
\section{The rest-frame 15 Micron Luminosity Functions}
\label{sec:15 LF}
\begin{table*}
\caption{\label{tab:15 k correc}15 microns \textit{k}-correction of the sample}
\centering
\begin{tabular}{cccccccccc}
\hline \hline
& \multicolumn{3}{c}{$0.4<z<0.7$} & \multicolumn{3}{c}{$0.7<z<1.0$}& \multicolumn{3}{c}{$1.0<z<1.3$}\\
& CE01 & LDP & DH & CE01 & LDP  & DH & CE01 & LDP & DH \\
\hline
Median & 1.29 & 1.16 & 1.16 & 0.88 & 0.90 & 0.89 & 0.98 & 0.95 & 0.97 \\
Mean & 1.32 & 1.15 & 1.15 & 0.87 & 0.91 & 0.90 & 1.00 & 0.97 & 0.99 \\
Sigma  & 0.20 & 0.06 & 0.07 & 0.11 & 0.08 & 0.07 & 0.15 & 0.10 & 0.11 \\
\hline
\end{tabular}
\end{table*}
\indent{
In this Section, we present the redshift evolution of the rest-frame 15\,$\mu$m luminosity function (LF) that we derive from the catalogs described in the previous section using the observed MIPS 24\,$\mu$m flux density.
The 15\,$\mu$m LF is computed for three redshift bins centered at z=0.55, 0.85 and 1.15 (same bins as in Figure~\ref{fig: optical}) for which the 24 $\mu$m observations correspond to the rest-frame 15, 13 and 11\,$\mu$m wavelengths respectively.
The extrapolation that needs to be applied to compute the 15 $\mu$m luminosity is therefore negligible in the first two redshift bins (still it is carefully computed here) and becomes larger in the highest redshift bin due to the fact that the observed 24 $\mu$m is more distant from the rest-frame 15 $\mu$m.
While the derivation of the rest-frame 15 $\mu$m LF has already been addressed in some previous studies \citep{lefloch_2005}, this is done here for two reasons. 
\\}
\indent{
First, we wish to check and demonstrate the consistency of our technique and catalogs with those previous studies at a common wavelength before deriving the rest-frame 35\,$\mu$m LF. As we show it here, our results are not only consistent with those previous studies but even extend them to fainter luminosities as well as they provide a more robust constraint on the 15\,$\mu$m LF because our catalogs are three times deeper at 24\,$\mu$m than the one used in \citet{lefloch_2005}. We are also able to validate the results against the effects of cosmic variance by comparing three independent fields: the EGS, the GOODS and the ECDFS-O fields.
\\}
\indent{
Second, this study will serve as a basis for the extension of the derivation of the faint end of the 35\,$\mu$m LF. For that purpose, we will first determine the typical ratio of the observed 70 over 24\,$\mu$m flux densities per redshift and flux density bins in order to associate 24\,$\mu$m sources to a given flux density bin at 70\,$\mu$m. Then we will derive the contribution of this bin to the 35\,$\mu$m LF.
\\}
\indent{
The sizes of the redshift bins [0.4-0.7], [0.7-1.0], [1.0-1.3] were defined such as to encompass a large enough number of sources in the computation of the 35\,$\mu$m LF. For consistency, we used the same redshift bins here for the 15\,$\mu$m LF.
\\}
\begin{figure}
	\includegraphics[width=8.5cm]{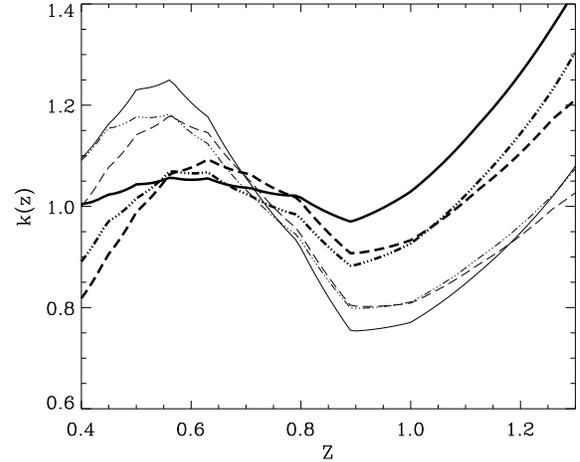}
	\caption{\label{fig:k-correc 15 micron}15 $\mu$m \textit{k}-correction as a function of the redshift (as in defined in Eq. \ref{eq:k correc}) for the MIPS-24 $\mu$m pass band.
	The thin and the thick solid lines correspond to the \textit{k}-correction obtained using the \citet{chary_2001} $10^{11}\,L_{\odot}$ and $10^{12}\,L_{\odot}$ templates respectively.
	The thin and the thick dashed lines correspond to the \textit{k}-correction obtained using the \citet{lagache_2003} $10^{11}\,L_{\odot}$ and $10^{12}\,L_{\odot}$ template respectively.
	Finally the thin and thick dashed-dot-dot lines correspond to the \textit{k}-correction obtained using the \citet{dale_2002} $10^{11}\,L_{\odot}$ and $10^{12}\,L_{\odot}$ template respectively.
	}
\end{figure}
\subsection{Methodology\label{subsec:methodo 15}}
\indent{
We define here the k-correction as the ratio between the rest-frame 15 and $24/(1+z)\,\mu$m luminosities as in Eq. \ref{eq:k correc}.
\begin{equation}\label{eq:k correc} 
\nu_{15\,\mu m} L_{\nu}^{15\,\mu m}=\nu_{24\,\mu m}4\pi d_{L}^{2}S(24\mu m)k(z)
\end{equation}
where $L_{\nu}^{15\,\mu m}$ is the rest-frame monochromatic luminosity in $\rm{W\,Hz^{-1}}$, S(24 $\mu$m) is the observed 24 $\mu$m flux density in $\rm{Wm^{-2}Hz^{-1}}$, $d_{L}$ is the luminosity distance and $k(z)$ is the $k$ correction.
This correction depends on the exact spectral energy distribution (SED) of the galaxies. In the following, we will use the Chary \& Elbaz (2001; hereafter CE01) library of 105 template SEDs as proxies of the typical SED of galaxies assuming that these SEDs, constrained by local galaxy properties, remain valid in the redshift range considered here. There are good reasons to believe that this is indeed the case, such as the consistency between the total IR luminosity derived from the observed mid-IR and radio flux densities of galaxies (Elbaz et al. 2002, Appleton et al. 2005, Marcillac et al. 2006). Until enough data are gathered combining broadband photometry and spectroscopy from IRS (the Spitzer infrared spectrograph), we use three available libraries of template SEDs to quantify the level of variations of the k-correction in this redshift range at first order. These three libraries are the CE01, LDP (Lagache, Dole \& Puget 2003) and DH \citep{dale_2002} ones. 
We convolve the redshifted SED templates with the \textit{Spitzer} 24 $\mu$m filter to compute the corresponding \textit{k}-corrections.
In Figure \ref{fig:k-correc 15 micron} we present the inferred \textit{k}-corrections for the three different libraries. 
We notice that for a given total infrared luminosity (i.e $L_{IR}=L(8-1000\,\mu\rm{m})$), all three libraries give a slightly different \textit{k}-correction as a function of redshift.
Nevertheless these discrepancies reach at most a value of 20$\%$. 
In Table \ref{tab:15 k correc} we present the average value of the \textit{k}-correction inferred for our sample in the three redshift bins for the three different libraries.
We notice that their average values are of the same order with a typical offset of 10\%, 5\% and 5\% in the three redshift bins respectively. Hence, the choice of a particular SED library should not affect much the computation of the rest-frame 15 microns LF. 
\\}
\indent{
In the following, the 15 microns luminosity ($\nu L_{\nu}^{15\mu m}$) of each galaxy is derived using the CE01 library.
This choice was made to be homogeneous with the rest of the study since we will also use the CE01 templates to derive the total infrared luminosity of galaxies.
Indeed, as illustrated in Section \ref{subsec:stacking}, the CE01 library appears to best reproduce the observed luminosity-luminosity correlations even at high redshift (i.e $z\thicksim1$). 
\\}
\begin{figure*}
\center{
	\includegraphics[width=17.5cm]{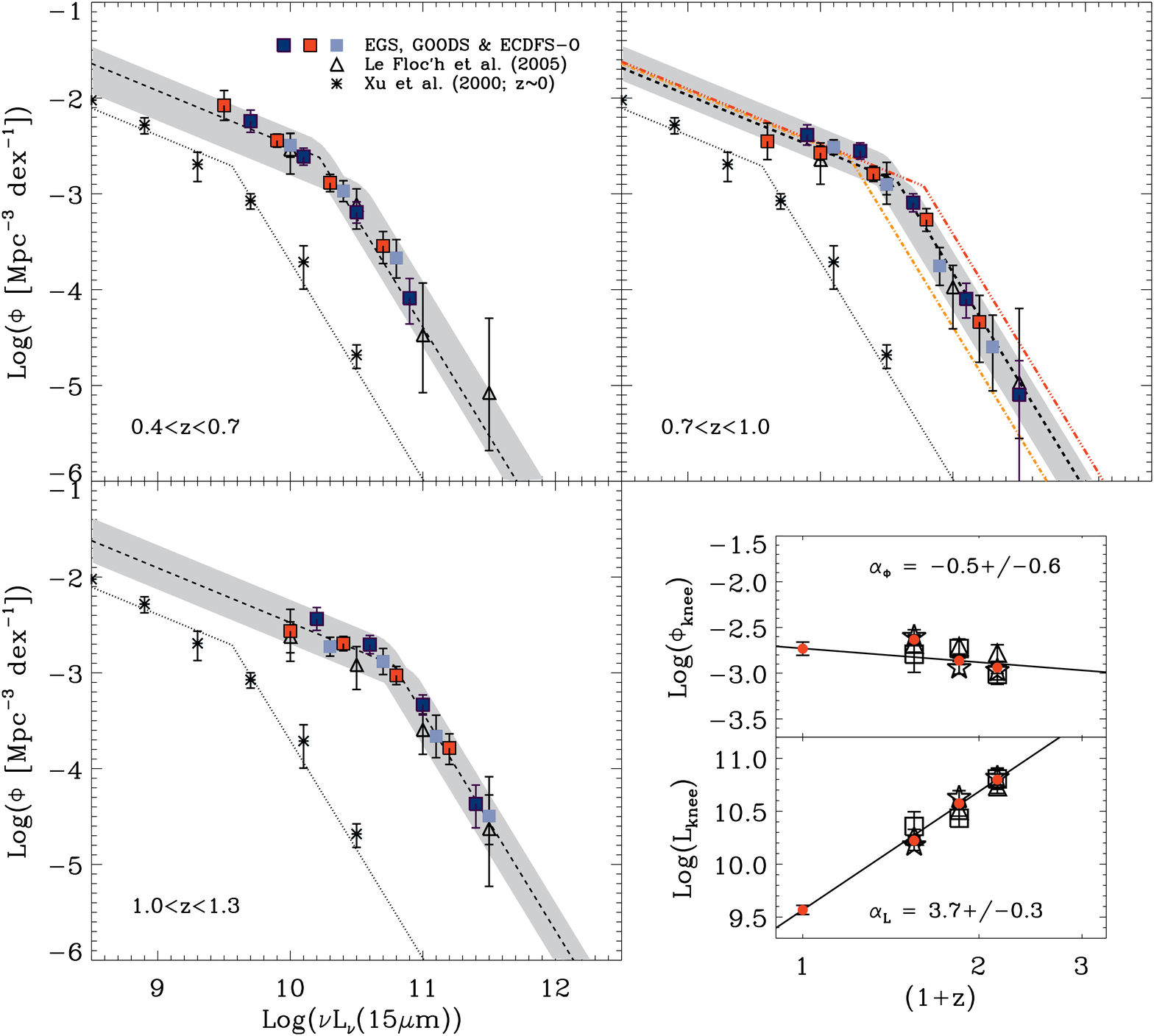}
	\caption{\label{fig:LF 15} Rest-frame 15 microns LF  estimated for three redshift bins with the $1/V_{max}$ method.
	The light red squares are obtained for the combined GOODS-North and South fields.
	The dark blue squares and the light blue squares are obtained for the EGS and the ECDFS-O fields respectively.
	Empty triangles show the LFs derived by \citet{lefloch_2005} in the ECDFS at redshifts $\thicksim0.7$, $\thicksim0.9$ and $\thicksim1.1$. 
	Asterisks show the local reference taken from \citet{xu_2000} and the dotted line is a fit to this data with a double power law function (see text).
	In the panel showing the $0.7<z<1.0$ redshift bin, we reproduced the best fit of the LF obtained in the two other redshift bins (i.e $0.4<z<0.7$ dot-dashed line and $1.0<z<1.3$ triple-dots-dashed line).
	The shade area in each panel spans all the parametric solutions obtained with the $\chi^{2}$ minimization method and compatible with the data within 1 $\sigma$.
	The inset plot represents the evolution of the $\phi_{knee}$ and $L_{knee}$ as function of redshift obtained from fits to all field (\textit{red fill circle}), GOODS (\textit{empty star}), EGS (\textit{empty triangle}) and ECDFS-O (\textit{empty square}).
	The redshift evolution of $L_{knee}$  and $\phi_{knee}$ was fitted using laws in the forms of $(1+z)^{\alpha_{L}}$ and $(1+z)^{\alpha_{\phi}}$, respectively.
	}}
\end{figure*}
\indent{
The LF are derived using the $1/V_{max}$ method \citep{schmidt_1968} which does not require any assumption on the shape of the LF.
The LF is then directly derived from the data.
The comoving volume associated to any source of a given luminosity is defined as $V_{max}=V_{zmax}-V_{zmin}$ where $zmin$ is the lower limit of the redshift bin, and $zmax$ is the maximum redshift at which the object could be seen given the flux density limit of the sample, with a maximum value corresponding to the upper redshift of the redshift bin.
For each luminosity bin, the LF is then given by
\\}
\begin{equation}\label{eq:sfr lir}  
	\phi = \frac{1}{\Delta L}\sum_{i}\frac{1}{V_{max,i}}w_{i}
\end{equation}
\indent{
where $V_{max}$ is the comoving volume over which the $i$th galaxy could be observed, $\Delta L$ is the size of the luminosity bin, and $w_{i}$ is the completeness correction factor of the $i$th galaxy.
$w_{i}$ equals 1 for sources brighter than $S_{24\mu m}\thicksim 100\,\mu$Jy and decreases at fainter flux densities due to the incompleteness of the 24 $\mu$m catalog (see the red histogram of the inset plot in Figure \ref{fig:simulation}). 
These completeness correction factors are robustly determined since they only affect the faintest luminosity bin and by at most a factor of 0.7. Hence none of the conclusions presented here strongly rely on this correction.
\\}
\indent{
The four different fields studied here (EGS, GOODS-N, GOODS-S and ECDFS-O) have different flux density limits, therefore we will treat separately the GOODS-N and S ($S_{24\mu m}^{limit}=30\,\mu$Jy), the EGS ($S_{24\mu m}^{limit}=50\,\mu$Jy) and the ECDFS-O ($S_{24\mu m}^{limit}=70\,\mu$Jy).
This will allow us to test the role of cosmic variance.
\\}
\indent{
The uncertainties of the LF values depend on the numbers of sources in each luminosity bin, on the photometric redshifts uncertainties, the \textit{k}-correction uncertainties and finally on the 24 and 70 $\mu$m flux uncertainties.
To compute these errors we performed Monte Carlo simulations by creating 1000 simulated catalogs.
These simulated catalogs contain the same number of sources as the original one but we attributed to each source with a photometric redshift a new redshift randomly selected within its probability distribution function.
Sources with a spectroscopic redshift have been left unchanged.
To take into account the redshift catastrophic errors we randomly selected $4\%$ of sources with a photometric redshift (corresponding to the fraction of catastrophic sources found in Section \ref{subsec: uncertainties}) and then attributed to these sources a random redshift selected assuming a flat distribution between $0<z<3$.
We notice that assuming a flat distribution may not be totally realistic since catastrophic errors are generally bimodal (e.g, assigning to a low redshift galaxy a much higher redshift due to mistaking one spectral break for another).
Nevertheless the sample that we use to compare the photometric and spectroscopic redshifts is not large enough to describe rigorously this bimodality.
However, assuming a flat distribution may be a more conservative assumption since this can yield an overestimation of the uncertainties due to redshift catastrophic errors.
Indeed if the bimodal distribution of the catastrophic errors is characterized by $\Delta z>3$ (i.e due to mistaking the Lyman break for a Balmer break) then it will only affect the estimates of the LF at $z>3$.
To take into account flux uncertainties we attributed to each source a new flux selected into a Gaussian distribution centered at the original flux of the source and with a dispersion computed as $\sigma=\sqrt{(\sigma_{calibration})^{2} + (\sigma_{source})^{2}}$ where $\sigma_{calibration}$ is the calibration uncertainty estimated to be $\thicksim10\%$ of the source flux density and $\sigma_{source}$ is the source photometric uncertainty computed on the residual image.
Finally to take into account the \textit{k}-correction uncertainties, we selected the source luminosity assuming a Gaussian distribution centered at the original source luminosity and with a dispersion of 0.05 dex.
Using these 1000 simulated catalogs, we then computed again the rest frame 15 $\mu$m LF and defined the total uncertainty in each luminosity bin as the quadratic sum of the Poissonian error ($\propto \sqrt{1/Nb_{sources}}$) and the dispersion given by the Monte-Carlo simulations.
We found that the Poisson statistics dominate the total uncertainties at the bright luminosity bin whereas at faint luminosities the total uncertainty is dominated by the Monte Carlo uncertainty.
\\}
\indent{
To study the evolution of the rest-frame 15 microns LF between $z\thicksim0$ and $z\thicksim1.3$ we used as a local reference the local 15 microns LF of \citet{xu_2000} derived using the bivariate (15 $\mu$m vs. 60 $\mu$m luminosity) method and based on ISOCAM observations.
This LF is then fitted with a double power law function.
This choice is driven by the fact that the ultimate goal of this paper is to derive the evolution of the total infrared LF which is itself best fitted with a double power law function \citep{sanders_2003}.
Hence to be homogeneous we have made the choice, in the following, to fit the 15, 35 $\mu$m and total infrared LFs with double power laws.
Using the \citet{sanders_2003} best-fit (i.e $\phi \propto L^{-0.6}$ for log($L/L_{\odot}$)$<10.5$ and  $\phi \propto L^{-2.15}$ for log($L/L_{\odot}$)$>10.5$) and a bivariate method, we derived the values of the two power laws for the 15\,$\mu$m LF.
We find that $\phi \propto L^{-0.5}$ for log($L/L_{\odot}$)$<\,L_{knee}$ and  $\phi \propto L^{-2.15}$ for log($L/L_{\odot}$)$>\,L_{knee}$.
Then, using a $\chi^{2}$ minimization, we fitted the \citet{xu_2000} observations with this function, leaving $L_{knee}$ and the normalization as free parameters.
We also fit independently, with the same method and the same function, the rest-frame 15 microns LF derived in the three redshift bins (see Table \ref{tab:fit parameter}).
\\}
\indent{
In Figure \ref{fig:LF 15} we present the rest-frame 15 $\mu$m LF derived with the $1/V_{max}$ method for the three redshift bins considered in this study.
As already mentioned, these LFs have been derived jointly for the GOODS-N and S fields and separately for the EGS and ECDFS-O fields.
The negligible role of cosmic variance is illustrated by the very good agreement between the three solutions in the derived 15 $\mu$m LF.
\\}
 \subsection{Results}
 \indent{
The resulting rest-frame 15 $\mu$m LFs are presented in Figure \ref{fig:LF 15}.
They provide three significant results.
 \\}
 \indent{
First, as written in the previous section, no strong effect of cosmic variance (i.e $\thicksim0.05-0.1$ dex) is found between the three fields, which provide the same 15\,$\mu$m LF within the error bars.
This observed estimate of the cosmic variance is totally consistent with theoretical estimates of $\thicksim0.07 $ dex based on the area of our fields  \citep[see][]{trenti_2008}.
 \\}
 \indent{
Second, the comparison of the rest-frame 15 $\mu$m LF obtained in this study and the ones inferred by \citet{lefloch_2005} are in agreement for all 3 redshift bins.
We note that the inferred LF was here extended to fainter luminosities using our $\thicksim3$ times deeper catalog which allowed us to obtain better constraints on the faint end slope of the LF.
\\}
\begin{table*}
\caption{\label{tab:35 correc}35 microns \textit{k}-correction of the sample}
\centering
\begin{tabular}{cccccccccc}
\hline \hline
& \multicolumn{3}{c}{$0.4<z<0.7$} & \multicolumn{3}{c}{$0.7<z<1.0$}& \multicolumn{3}{c}{$1.0<z<1.3$}\\
&{CE01} &{LDP} &{DH} & {CE01} & {LDP}  &{DH} & {CE01} & {LDP} & {DH} \\
\hline
Median & 0.71 & 0.62 & 0.77 & 0.87 & 0.81 & 0.89 & 1.12 & 1.14 & 1.14 \\
Mean & 0.73 & 0.64 & 0.77 & 0.88 & 0.82 & 0.91 & 1.12 & 1.15 & 1.15 \\
Sigma  & 0.05 & 0.05 & 0.04 & 0.05 & 0.06 & 0.06 & 0.05 & 0.07 & 0.04 \\
\hline
\end{tabular}
\end{table*}
 \indent{
Third, we find a substantial evolution of the LF with redshift.
 This evolution, already revealed by previous studies \citep{lefloch_2005,chary_2001,caputi_2007,xu_2000,lagache_2003,franceschini_2001} is usually expressed through an evolution both in density and in luminosity of the local LF (i.e $\rho(L,0)$).
Assuming that the shape of the LF remains the same since $z\thicksim0$, these studies express the evolution  as $\rho(L,z)=g(z)\rho(L/f(z),0)$, where $g(z)$ and $f(z)$ describe the density and the luminosity evolution through $g(z)=(1+z)^{p}$ and $f(z)=(1+z)^{q}$.
The weakness of this parameterization is that it assumes a monotonic evolution of the LF preventing any turn-over in the evolution (e.g an increase in density that would be followed by a decrease).
This kind of turn-over has already been seen at other infrared wavelengths \citep[e.g][]{caputi_2007}.
Therefore, we chose to fit independently the 3 different redshift bins with double power laws.
Then tracking the value of $L_{knee}$ and $\phi_{knee}$ (see Sect. \ref{subsec:methodo 15}) as a function of redshift enabled us to constrain the evolution of the LF.
We find that the variation of $\phi_{knee}$ with redshift is not large and starts with a small rise from $z\thicksim0$ to $z\thicksim0.55$ followed by a small fall up to $z\thicksim1.15$.
Hence the 15 $\mu$m LF appears to be rather consistent with the standard definition of a pure, monotonic, luminosity evolution in this redshift range, proportional to $f(z)=(1+z)^{3.7\pm0.3}$.
We note that the variations in density or of $\phi_{knee}$ are small and of the same order than the difference in the galaxy number densities found between the GOODS and EGS fields in the faint end of the LF. Such small variations could well be produced by cosmic variance.
 \\}
\section{The rest-frame 35 Microns Luminosity Functions}
\label{section: function 35}
\indent{
We now compute the rest-frame 35 $\mu$m LF, based on 70 $\mu$m observations, to minimize \textit{k}-correction in the $z<1.2$ range and hence reduce the dependency of the LF on the assumed SED library.
\\}
 \begin{figure}
	\includegraphics[width=8.5cm]{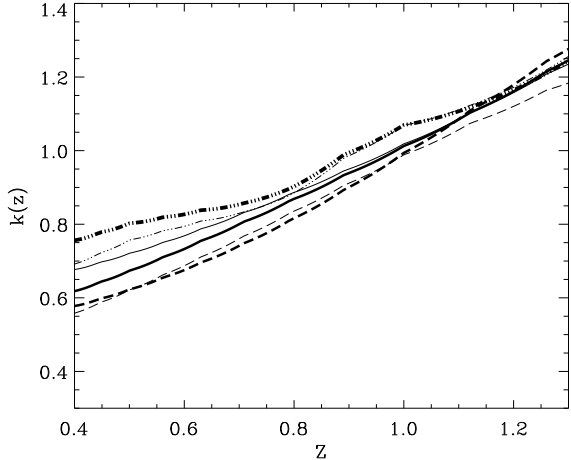}
	\caption{\label{fig:k-correc 35 micron}35 $\mu$m \textit{k}-correction as a function of the redshift for the MIPS-70 $\mu$m pass band.
	Lines are as in Figure \ref{fig:k-correc 15 micron}.
		}
\end{figure}
\subsection{Methodology\label{subsec:methodo 35}}
\indent{
In order to minimize \textit{k}-correction uncertainties due to potentially evolving SED with redshift, we chose to derive the FIR LF at the closest rest-frame wavelength associated to the MIPS 70 $\mu$m filter in the two most distant redshift bins (i.e $0.7<z<1.0$, $1.0<z<1.3$), which translate into the rest-frame 30-40 $\mu$m wavelength range. Therefore we chose to derive the rest-frame LF at  35 microns.
In the lowest redshift bin (i.e $0.4<z<0.7$), the MIPS 70 $\mu$m filter probes the rest-frame 40-50 $\mu$m wavelength range where the \textit{k}-correction is larger but local SEDs such as those used here were proved to be quite robust at these low redshifts \citep{bavouzet_2008}.
\\}
\indent{
As already discussed in this paper, the three different SED libraries give \textit{k}-corrections which agree within 10\% on average and up to 20\% at most around $z\thicksim0.5$ (see Figure \ref{fig:k-correc 35 micron} and Table \ref{tab:35 correc}). In the following we use the CE01 library and we will show in Section \ref{subsec:stacking} that this is indeed the best one, among the three considered here, for our purpose.
\\ \\}
\indent{
The rest-frame 35 $\mu$m LF was derived using the $1/V_{max}$ method and the \textit{k}-corrected 70 $\mu$m data.
The completeness correction factors were derived using the simulations described in Section \ref{subsec: prior method}.
The GOODS-S and N fields are treated jointly (flux limit $\thicksim 2.5$ mJy), the EGS and the ECDFS-O were treated separately (flux limit $\thicksim 3$ mJy and $\thicksim 3.5$ mJy).
As for the 15 microns LF, to take into account most of the uncertainties, we performed Monte Carlo simulations.
For these Monte Carlo simulations the 35 $\mu$m \textit{k}-correction uncertainties were assumed to be of 0.08 dex. This conservative value was derived using the largest disagreement observed between the three SED libraries. Indeed, as shown in Figure~\ref{fig:k-correc 35 micron}, we observe at z~0.6 a disagreement of ~0.08 dex between the LDP and DH libraries.
\\}
\indent{
There is no direct local reference available in the literature for the 35 $\mu$m LF.
Therefore we constructed a local reference from the rest-frame 25 microns LF of \citet{shupe_1998} derived using the IRAS Faint Source Survey.
Using a bivariate method and the CE01 library, we converted the 25 microns LF into a rest-frame 35 microns LF. 
As already mentioned, in the sake of consistency all along this study, we fitted this local reference and the three other LFs (corresponding to our three redshift bins) with double power laws with fixed slopes, i.e $\phi \propto L^{-0.6}$ for log($L/L_{\odot}$)$<\,L_{knee}$ and  $\phi \propto L^{-2.2}$ for log($L/L_{\odot}$)$>\,L_{knee}$ (see Table \ref{tab:fit parameter}).
\\}
\indent{
We have computed the rest-frame 35\,$\mu$m LF in two successive steps.
As discussed above, we started with the same technique as for the rest-frame 15\,$\mu$m LF, using the catalog of sources detected at 70\,$\mu$m and correcting for incompleteness, k-correction and 1/Vmax.
The results are presented in the Figure \ref{fig:LF 35} with filled stars for the EGS, GOODS (N \& S) and ECDFS-O fields.
For each redshift bin, we fitted the rest-frame 35\,$\mu$m LF using the local double power law reference with fixed slopes.
In Figure \ref{fig:LF 35} the light shaded areas represent the areas spanned by all the parametric solutions which are compatible within 1 $\sigma$ with the rest-frame 35\,$\mu$m derived using the 70 $\mu$m observations. 
Here again, we find that the bright end of the LF is well constrained and does not show any evidence for a change of slope.
The regime that we were able to probe down to the 70\,$\mu$m detection limit of 2 mJy corresponds to total IR luminosities of 0.7, 2 and 5$\times$10$^{11}$ L$_{\sun}$ computed from the CE01 library.
Hence, apart from the upper redshift bin, this luminosity domain covers the full range of LIRGs and ULIRGs.
This brings direct constraints, at these redshifts, to the relative contributions of LIRGs and ULIRGs to the comoving SFR density. Previous studies, based either on extrapolations from the mid-IR or on uncertain dust extinction corrections from the UV, were likely less robust than the derivation of a total IR luminosity from rest-frame 35\,$\mu$m luminosities.
In a second step, presented in the next section, we extend our derivation of the 35\,$\mu$m LF below its knee by using a stacking analysis.
\\}
\subsection{Constraining the faint end 35 microns LF through stacking\label{subsec:stacking}}
\begin{figure*}
\center{
	\includegraphics[width=17.5cm]{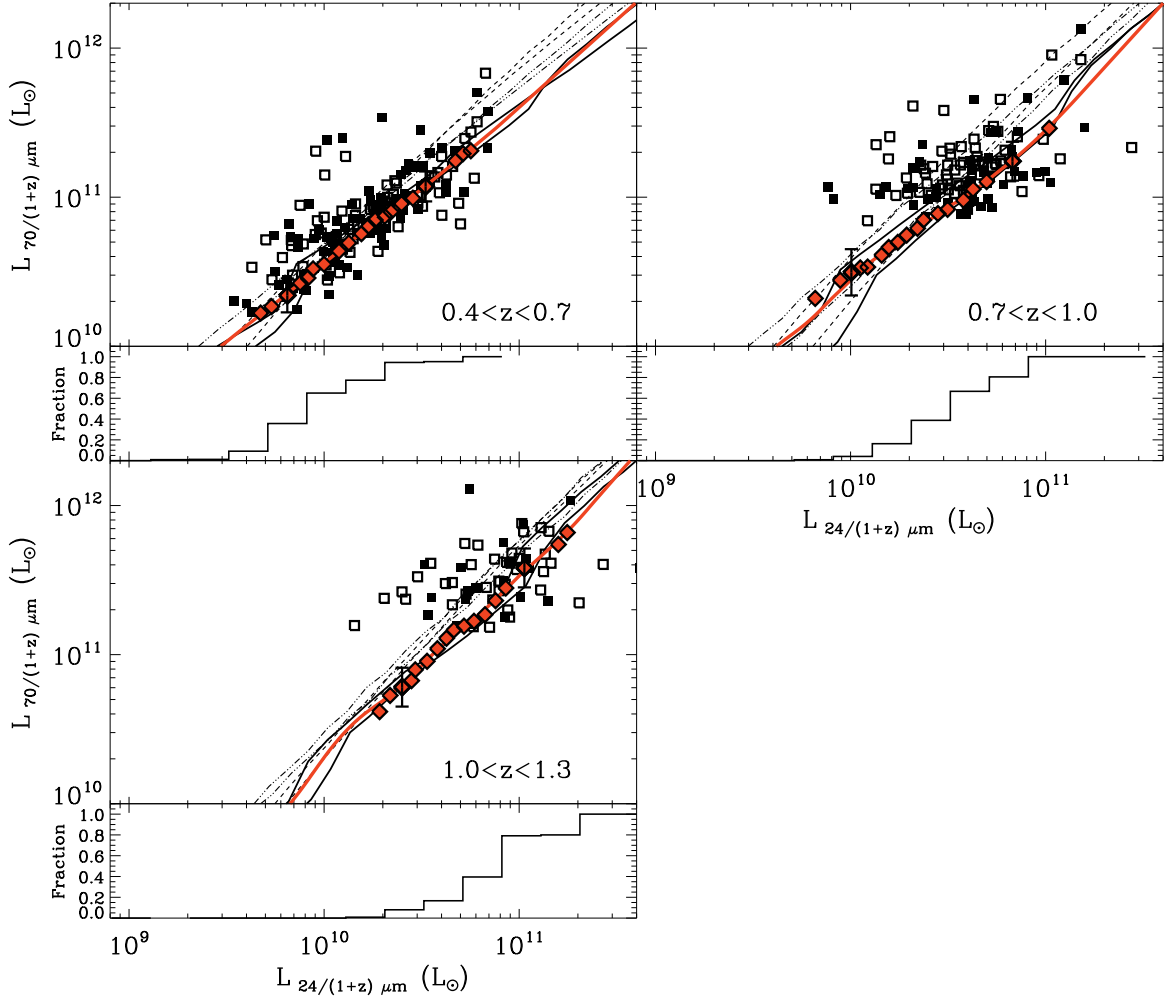}
	\caption{\label{fig:stacking}The 70 Vs. 24 $\mu$m correlations as revealed by the observations and our stacking analysis in the three redshift bins considered in this study.
	The empty and filled square present the 70 Vs. 24 $\mu$m correlations observed for sources which have a photometric and a spectroscopic redshift respectively.
	The red diamond show the results obtained using our stacking analysis (see text).
	The solid lines, the dashed lines and the triple-dots-dash lines represent the expected correlations for the CE01, the LDP and the DH library, respectively, at the lowest and the highest redshift of each redshift bin.
	The red solid line represents the inferred 24/70 $\mu$m correlation.
	At the bottom of each plot we present the fraction of 24 $\mu$m sources detected at 70 $\mu$m as a function of the 24/(1+z) $\mu$m luminosity.
	}
}
\end{figure*}
\begin{figure*}
\center{
	\includegraphics[width=17.5cm]{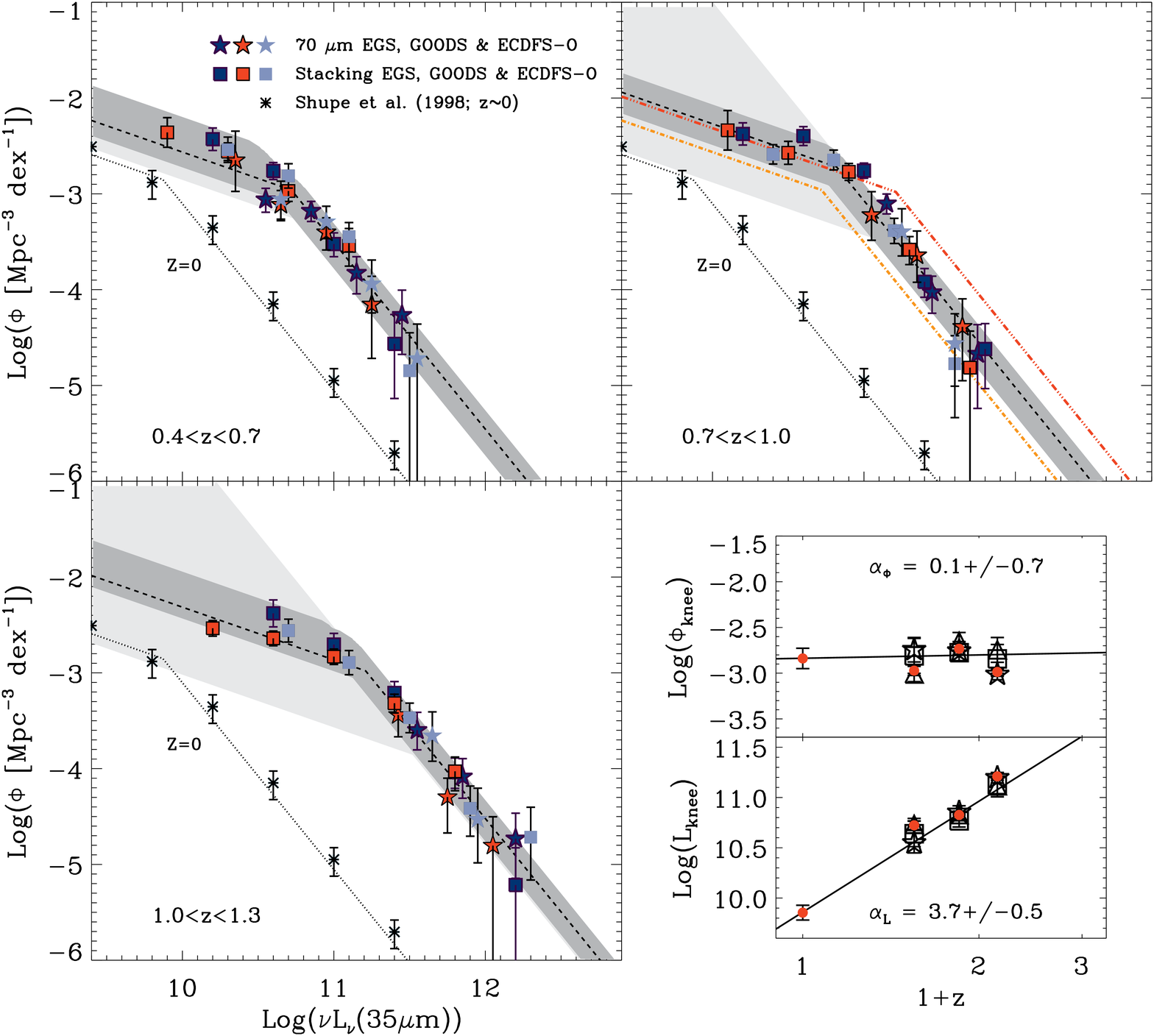}
	\caption{\label{fig:LF 35}Rest-frame 35 microns LF  estimated independently in three redshift bins with the $1/V_{max}$ method.
	The red stars are obtained from the GOODS-North and South fields using 70 $\mu$m sources.
	The dark blue and light blue stars show the LF obtained from 70 $\mu$m sources in  the EGS and the ECDFS-O fields, respectively.
	The red and dark blue squares are obtained starting from 24 $\mu$m and using the 24--70\,$\mu$m correlation of Figure~\ref{fig:stacking} in the GOODS and EGS fields, respectively.
	Asterisks show the local reference taken from \citet{shupe_1998} and the dotted line presents the best-fit to these data points with a double power law function (see text).
	In the intermediate redshift panel, we reproduced the best fit of the LF obtained in the two other redshift bins (i.e $0.4<z<0.7$ dot-dash line and $1.0<z<1.3$ triple-dots-dash line).
	The light and dark shaded area span all the solutions obtained with the $\chi^{2}$ minimization method and compatible, within 1 $\sigma$, with the LF measured using direct 70 $\mu$m observations and the LF measured using the stacking analysis, respectively.
	The inset plot represents the evolution of the $\phi_{knee}$ and $L_{knee}$ as function of redshift.
	Symbols and lines are as in Figure \ref{fig:LF 15}.
	}
}
\end{figure*}
\indent{
In order to stack positions of 35\,$\mu$m sources below the detection limit, it is necessary to adopt some prior positions.
The 24\,$\mu$m catalogs provide the best proxy for this purpose since the 24\,$\mu$m images are two orders of magnitudes deeper than the 35\,$\mu$m ones and the two wavelengths are the closest available.
If the IR SED of galaxies of a given luminosity would not vary much since $z\sim$1, we would expect a direct correlation between the 24\,$\mu$m/(1+$z$) and 70\,$\mu$m/(1+$z$) luminosities since local galaxies do follow tight correlations between their 6.75, 12, 15, 25, 60 and 100\,$\mu$m luminosities (see e.g. Chary \& Elbaz 2001).
We have tested the existence of such correlations both in the bright luminosity range where galaxies are detected at 24 and 70\,$\mu$m and in the lower luminosity range by stacking 70\,$\mu$m images on the positions of sources detected only at 24\,$\mu$m.  
\\}
\indent{
We started by dividing the 24\,$\mu$m catalogs into three sub-samples, one for each redshift bin. 
Then, we separated 24\,$\mu$m sources per luminosity bins of 0.4 dex. 
For each 24 $\mu$m luminosity bin we stacked on the residual 70\,$\mu$m image all sources with no 70 $\mu$m detection.
The photometry of the stacked image was then measured using an aperture of 16\arcsec, a background within annuli of 40 and 60\arcsec and an aperture correction factor of 1.705 (as discussed in the {\it Spitzer} observer's manual).
Finally, the mean 70\,$\mu$m flux density ($F^{70\,\mu m}_{bin}$) for a given 24 $\mu$m bin was computed following Equation \ref{eq:mean stack}.
\begin{center}
\begin{equation}\label{eq:mean stack} 
F^{70\,\mu m}_{bin}=\frac{m\times F^{70\,\mu m}_{stack}+\sum_{i=1}^{n}F^{70\,\mu m}_{i}}{n+m}
\end{equation}
\end{center}
where $F^{70\,\mu m}_{stack}$ is the 70 $\mu$m flux density, obtained through stacking, of all the 24 $\mu$m sources undetected at 70 $\mu$m (sample which contains $m$ sources) and $F^{70\,\mu m}_{i}$ is the 70 $\mu$m flux density of the $i$th 24 $\mu$m sources detected at 70 $\mu$m (sample which contains $n$ sources).
This procedure was performed using a sliding 24 $\mu$m luminosity bin and a step of 0.1 dex.
The result of this stacking analysis is shown with filled red diamonds in Figure \ref{fig:stacking}.
We notice that for the brightest 24 $\mu$m luminosity bins, $F^{70\,\mu m}_{bin}$  is totally dominated by the flux of detected sources since only a small fraction of the sources are not detected at 70 $\mu$m.
On contrary for the faintest 24 $\mu$m luminosity bins, $F^{70\,\mu m}_{bin}$ is dominated by $F^{70\,\mu m}_{stack}$.
\\}
\indent{
Uncertainties of stacking values depend on the homogeneity of the underlying sample and on background fluctuations.
To estimate the uncertainties related to the sample inhomogeneity (i.e $\sigma_{inhomo}$) we perform a standard bootstrap analysis.
To estimate the uncertainties due to background fluctuations (i.e $\sigma_{back}$) we perform a stacking analysis at random positions using the same numbers of stacked elements.
This stacking analysis at random positions is repeated 100 times and then $\sigma_{back}$ is given by the dispersion of these stacking fluxes. 
Finally the total uncertainty is defined as the quadratic sum of $\sigma_{inhomo}$ and $\sigma_{back}$.
These uncertainties are shown in Figures \ref{fig:stacking}.
\\}
\indent{
The first result that we find here is that in both regimes - i.e. where sources are strong enough to be detected at both wavelengths and where sources are only seen at 24\,$\mu$m - the two luminosities in the rest-frame wavelengths 24\,$\mu$m/(1+$z$) and 70\,$\mu$m/(1+$z$) do follow a correlation.
This correlation is in agreement  with the CE01 library but is a factor of two below predictions from the LDP and DH libraries.
This correlation being found in the three redshift bins it confirms that one can complement the 70\,$\mu$m LF at the faint end based on 24\,$\mu$m detections.
Therefore using the inferred 24/70 $\mu$m correlation (red line in Figure \ref{fig:stacking}) we have derived the 70\,$\mu$m luminosity of each source starting from their 24\,$\mu$m flux density and then produced the full 35\,$\mu$m LFs.
These LFs are represented with filled squares in Figure \ref{fig:LF 35}.
Using the local double power law reference with fixed slopes, we then fitted these rest-frame 35\,$\mu$m LFs and all solutions of this fit compatible, within 1 $\sigma$, with the data are presented as the dark shaded area of Figure \ref{fig:LF 35}.
As revealed by the overlapping of the light and dark shaded areas, the resulting 35\,$\mu$m LF are fully consistent with the ones based on direct 70\,$\mu$m detections and extend them far below the knee of the LF.
\\}
  \subsection{Results}
\indent{
We used the combination of direct 70\,$\mu$m detections and 24\,$\mu$m-derived luminosities, converted into 35\,$\mu$m luminosities from stacked 70\,$\mu$m images, to derive 35\,$\mu$m LF over more than two orders of magnitude, allowing us to position the knee of the LF in all three redshift bins (see Figure \ref{fig:LF 35}). The data do not show any obvious sign for a change of slope of the LF both at the faint and bright ends, from $z\sim$0 to 1.3. This confirms that our choice to keep the same double power-law slopes than the ones used locally is pertinent. Like for the 15\,$\mu$m LF, we only find marginal evidence for a variation of $\phi_{knee}$ with redshift, while $L_{knee}$ follows a rapid evolution proportional to $(1+z)^{3.6\pm0.5}$ law from $z\thicksim0$ to $z\thicksim1.3$, consistent with pure luminosity evolution. The flat slope of the bright end of the LF, typical of the IR LF, and its rapid evolution with redshift clearly illustrates the rising power of luminous galaxies over less luminous ones when going to higher redshifts. In the following section, we discuss the implications of the redshift evolution of the 35\,$\mu$m LF on the total IR LF.
\\}
\begin{figure}
	\includegraphics[width=8.5cm]{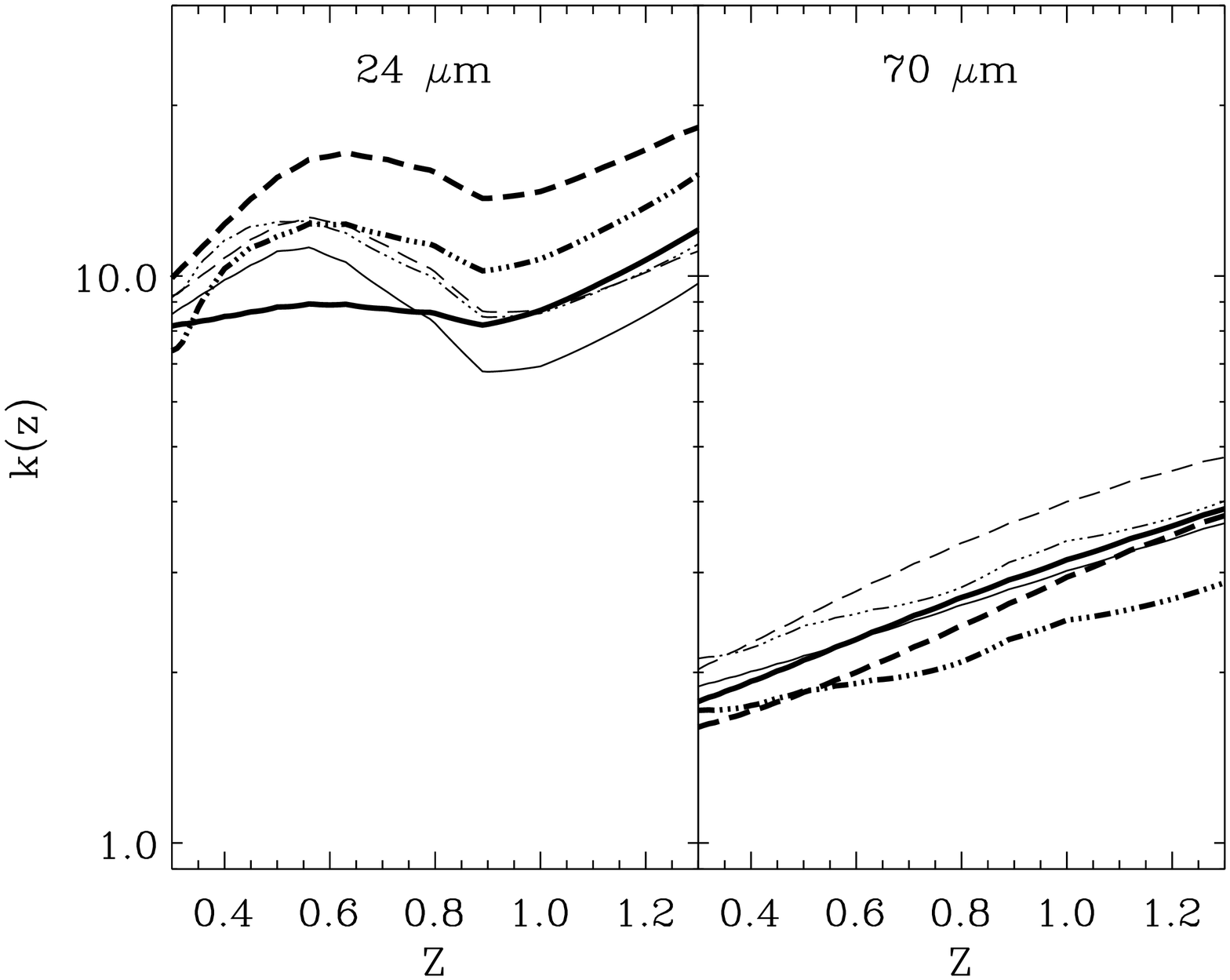}
	\caption{\label{fig:lir conversion}The \textit{k}-correction between the observed 24 $\mu$m luminosity and the total infrared luminosity as a function of redshift (\textit{left}) and the \textit{k}-correction between the observed 70 $\mu$m luminosity and the total infrared luminosity as a function of redshift (\textit{right}).
	Lines are as in Figure \ref{fig:k-correc 15 micron}.
}
\end{figure}
\section{Total Infrared Luminosity Function}
\label{sec:ir LF}
\indent{
In this section, we use the best proxy to compute the total IR luminosity of galaxies (L$_{\rm IR}$), i.e. the observed 70\,$\mu$m data when available or else the 24\,$\mu$m data. L$_{\rm IR}$ is computed over the wavelength range 8--1000\,$\mu$m. Under this definition, the SFR and L$_{\rm IR}$ are related by Eq.~\ref{eq:SFR_Lir} (Kennicutt 1998) for a Salpeter IMF ($\phi$(m)$\propto$ m$^{-2.35}$, between 0.1--100 M$_{\odot}$). Note that we integrate the SED curve to derive L$_{\rm IR}$ instead of using the classical approximation using the four IRAS broadband filters as defined in \citet{sanders_1996}.
\begin{equation}\label{eq:SFR_Lir} 
SFR~ [{\rm M}_{\odot}~ {\rm yr}^{-1}] = 1.72 \times 10^{-10} L_{\rm IR} ~[{\rm L}_{\odot}]
\end{equation}
Up to a redshift of $z\thicksim1$, the local $\nu L_{\nu}$-$L_{\rm IR}$ correlations appear to remain valid (see Bavouzet et al. 2008, Elbaz et al. 2002, Appleton et al. 2004) and the wavelengths closest to the peak far IR emission provide the most accurate estimators of L$_{\rm IR}$ (Bavouzet et al. 2008). The k-correction factors required to convert a measurement at 70 $\mu$m into an L$_{\rm IR}$ present a smooth evolution with redshift, consistently followed by all three SED libraries discussed in the previous sections (see Fig.\ref{fig:lir conversion}-right). However, when using the 24 $\mu$m band, the three libraries vary by a factor of $\thicksim0.3$ dex. Hence using 70 $\mu$m data provide an estimate of L$_{\rm IR}$ nearly independent of the SED library used.
We have seen in the previous section that the CE01 library of template SEDs provide a good estimate of the 70\,$\mu$m emission of galaxies based on observed 24\,$\mu$m data. Therefore, we will use that same library to then extrapolate from observed 70\,$\mu$m to L$_{\rm IR}$.
\\}
\begin{figure*}
	\includegraphics[width=17.5cm]{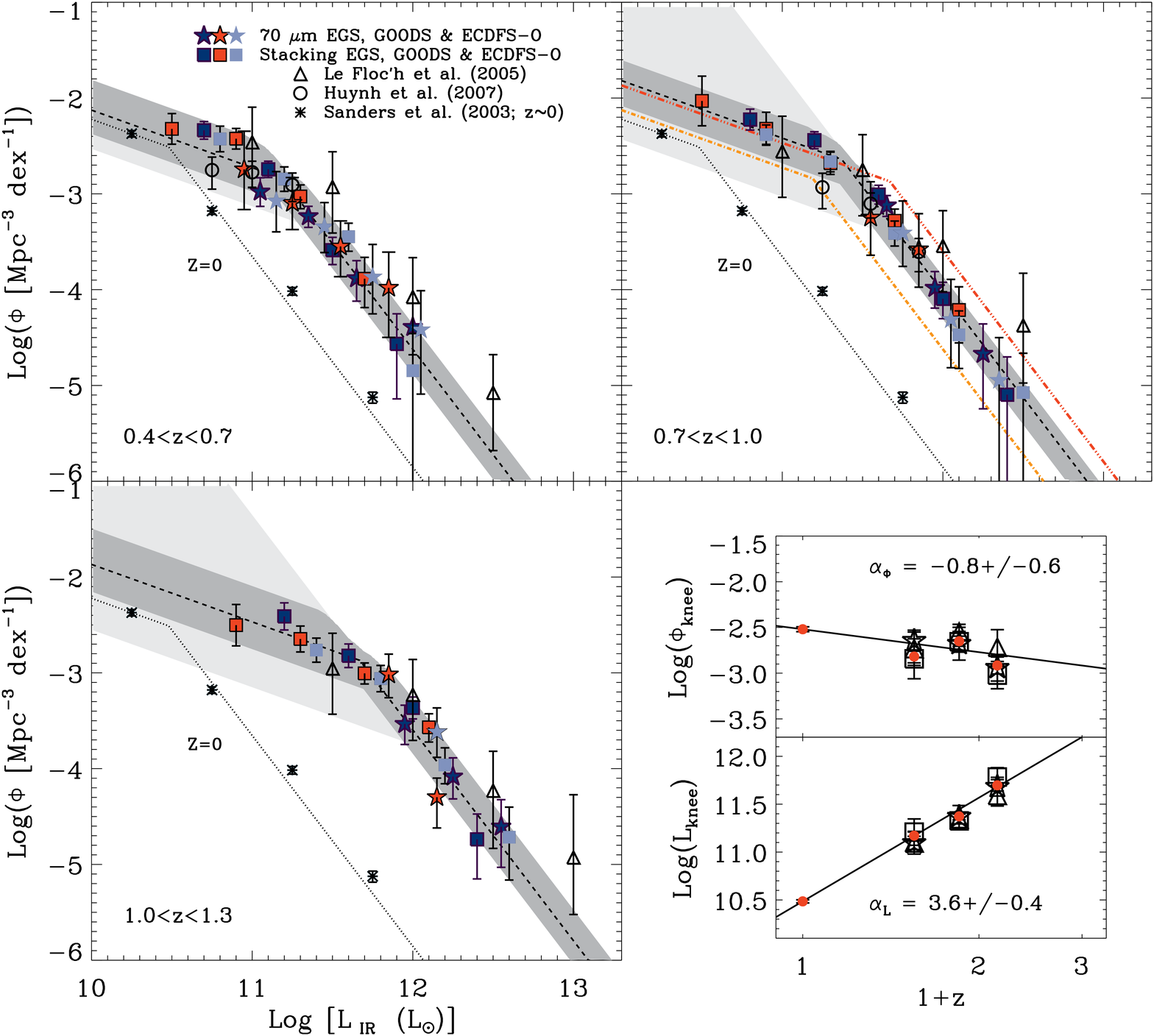}
	\caption{\label{fig:LF IR}Total infrared LF estimated for three redshift bins with the $1/V_{max}$ method.
	Lines and symbols are the same as in Figure \ref{fig:LF 35}.
	Asterisks show the local reference taken from \citet{sanders_2003} and the dotted line presents the fit of these data points with a double power law (see text).
	The empty triangles and the empty circles are taken from \citet{lefloch_2005} and \citet{huynh_2007}, respectively.
	The light and dark shaded area spans all the parametric solutions obtained with the $\chi^{2}$ minimization method and compatible, within 1 $\sigma$, with the LF measured using direct 70 $\mu$m observations and the LF measured using the stacking analysis respectively.
	The inset plot represents the evolution of the $\phi_{knee}$ and $L_{knee}$ as function of redshift.
	Symbols and lines are the same as in Figure \ref{fig:LF 15}.
}
\end{figure*}
\subsection{Luminosity function}
\begin{table*}
\caption{\label{tab:fit parameter}Parameter values of the 15 $\mu m$, 35 $\mu m$ and total infrared LF}
\centering
\begin{tabular}{cccccc}
\hline \hline
{Redshift} & {Wavelength} &{$\alpha_{1}\,^\mathrm{a}$} &{$\alpha_{2}\,^\mathrm{a}$} &{\rm{Log($L_{knee}$)}$\,^\mathrm{b}$ } & {\rm{Log($\phi_{knee}$)}$\,^\mathrm{b}$} \\
& & & &  {\tiny{\rm{Log($L_{\odot}$)}}} & {\tiny{\rm{Log($Mpc^{-3}dex^{-1}$)}}} \\
\hline
$z\thicksim0$ & 15 $\mu$m & $-0.57$ & $-2.27$ &  $9.56\pm0.04 $ & $-2.73\pm0.07$\\
$0.4<z<0.7$ & 15 $\mu$m & $-0.57$ & $-2.27$ & $10.22\pm0.03$ & $-2.63\pm0.05$\\
$0.7<z<0.7$ & 15 $\mu$m & $-0.57$ & $-2.27$ & $10.57\pm0.04$ & $-2.86\pm0.04$\\ 
$1.0<z<0.7$ & 15 $\mu$m & $-0.57$ & $-2.27$ & $10.79\pm0.05 $ & $-2.93\pm0.06$\\ \\ 
$z\thicksim0$ & 35 $\mu$m & $-0.55$ & $-1.95$ & $9.85\pm0.07 $ & $-2.83\pm0.10$\\
$0.4<z<0.7$ & 35 $\mu$m & $-0.55$ & $-1.95$ & $10.73\pm0.04$ & $-2.98\pm0.05$\\
$0.7<z<1.0$ & 35 $\mu$m & $-0.55$ & $-1.95$ & $10.82\pm0.04$ & $-2.73\pm0.06$\\
$1.0<z<1.3$ & 35 $\mu$m & $-0.55$ & $-1.95$ & $11.20\pm0.04 $ & $-2.98\pm0.05$\\ \\
$z\thicksim0$ & IR & $-0.60$ & $-2.20$ & $10.48\pm0.02 $ & $-2.52\pm0.03$\\
$0.4<z<0.7$ & IR & $-0.60$ & $-2.20$ & $11.19\pm0.04$ & $-2.84\pm0.06$\\
$0.7<z<1.0$ & IR & $-0.60$ & $-2.20$ & $11.37\pm0.03$ & $-2.65\pm0.05$\\
$1.0<z<1.3$ & IR & $-0.60$ & $-2.20$ & $11.69\pm0.06 $ & $-2.91\pm0.10$\\
\hline
\end{tabular}
\begin{list}{}{}
\item[$^{\mathrm{a}}$] Fixed slopes
\item[$^{\mathrm{b}}$] These values are inferred using direct 70 $\mu$m observation and the stacking analysis
\end{list}
\end{table*}
 \indent{
The total infrared luminosity function is derived using the $1/V_{max}$ method and the completeness correction factor derived from our simulations.
The two GOODS fields were combined and then the EGS, ECDFS-O and GOODS fields were treated separately.
Error bars were derived using Monte Carlo simulations to take into account most of the uncertainties.
For these Monte Carlo simulations the total infrared \textit{k}-correction uncertainties were assumed to be of 0.11 dex.
\\}
\indent{
We use as a local reference for the total IR LF, the \citet{sanders_2003} LF derived from the analytical fit to the IRAS Revised Bright Galaxy Sample, i.e $\phi \propto L^{-0.6}$ for log($L/L_{\odot}$)$<\,L_{knee}$ and  $\phi \propto L^{-2.2}$ for log($L/L_{\odot}$)$>\,L_{knee}$ with $log_{10}(L_{knee}/{\rm L}_{\odot}$)=10.5 ($z\sim0$).
\\}
\indent{
As for the rest-frame 35 $\mu$m LF, we use the observed 24 $\mu$m to constrain the faint end slope of the LF.
For each source we first derived its 70 $\mu$m flux density using the 24-70 $\mu$m correlations that we obtained from our stacking analysis (see Section \ref{subsec:stacking}).
Then, as for direct 70 $\mu$m detections, we used the CE01 SEDs to derive L$_{\rm IR}$ from the 70 $\mu$m flux density.
\\}
\indent{
We separate the values of the LF derived from observed 70 $\mu$m (stars) and stacked 70\,$\mu$m (squares) in the resulting total IR LF (Figure \ref{fig:LF IR}). 
Using the \citet{sanders_2003} local reference with fixed slopes we fit independently the total infrared LF derived using direct 70 $\mu$m observations (light shaded area) and the one derived using the stacking analysis (dark shaded area). 
We note that direct 70 $\mu$m observations allow us to sharply constrain the evolution of LIRGs and ULIRGs up to $z\thicksim1.15$ but fail to constrain the knee of the total IR LF which requires to rely on stacking (see light shaded area). In the regime where direct 70 $\mu$m data is available, we find that the two techniques (starting from either 24 $\mu$m $+$ stacking or 70 $\mu$m) provide consistent bright sides of the LF. The resulting parametric fits (using the parametric form of the local reference at all redshifts) are summarized in Table \ref{tab:fit parameter}.
\\}
\indent{
The total IR LF is found to exhibit a rapid evolution with redshift (Fig.~\ref{fig:LF IR}) nearly consistent with a pure luminosity evolution proportional to $(1+z)^{3.6\pm0.4}$ from $z\thicksim0$ to $z\thicksim1.3$. Marginal evidence for a decrease of $\phi_{knee}$ is found proportional to $(1+z)^{-0.8\pm0.6}$ when fitting all fields, but staying consistent with a flat behavior when accounting for cosmic variance (see Fig.~\ref{fig:LF IR}).
\\}
\begin{figure}
	\includegraphics[width=8.5cm]{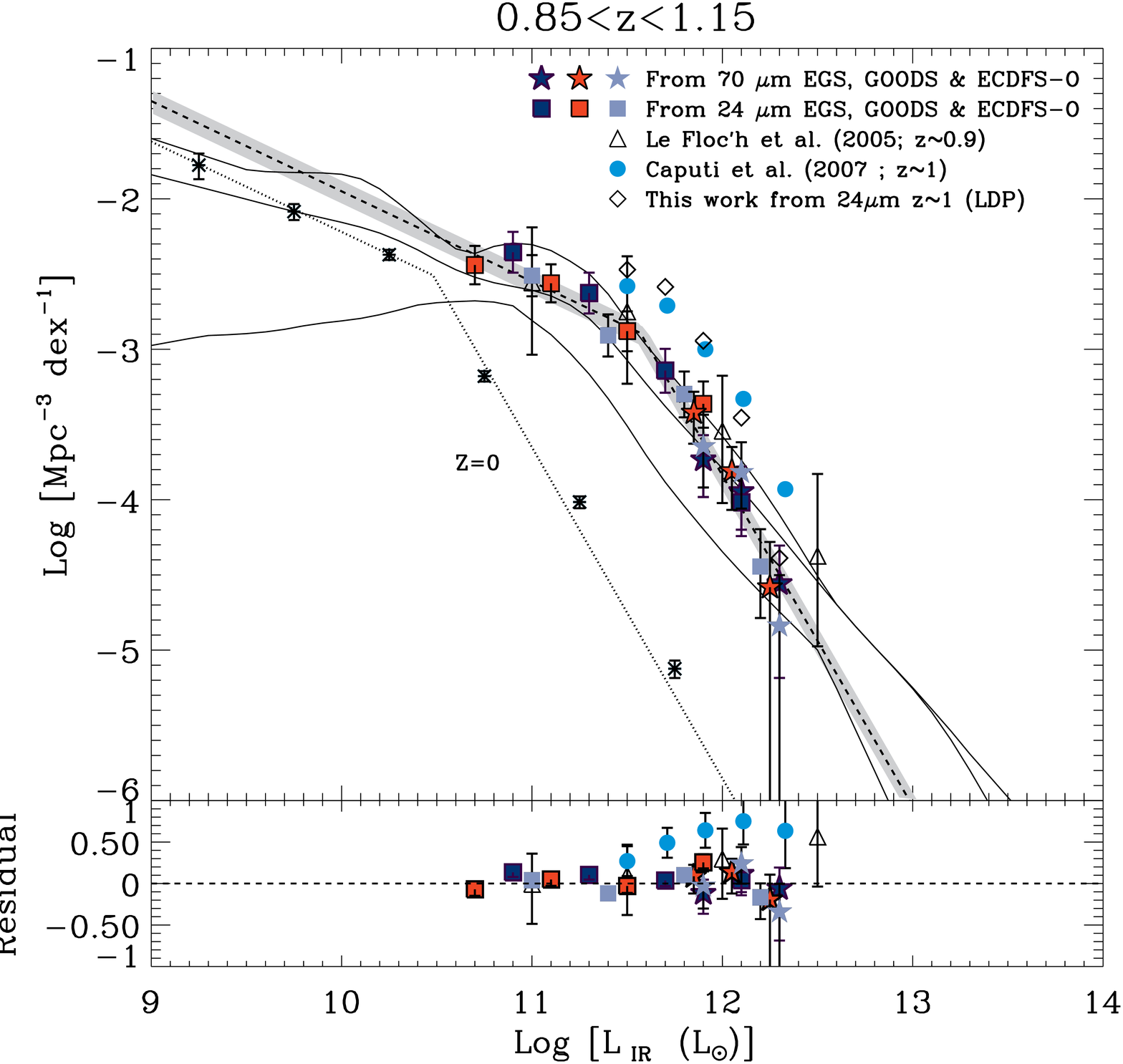}	
	\caption{\label{fig:lir caputi}Total infrared LF estimated at $z\thicksim1$ with the $1/V_{max}$ method.
	Symbols are the same as in Figure \ref{fig:LF IR} except for the blue circles which represent the LF inferred by \citet{caputi_2007} at $z\thicksim1$ using 24 $\mu$m observations and the empty diamonds which represent the LF that we would have inferred using our 24 $\mu$m observations and the \citet{lagache_2003} SED library to convert monochromatic observations into total infrared luminosity.
	The shaded area represents the surface spanned by all the parametric solutions obtained from the $\chi^{2}$ minimization method and compatible within 1 $\sigma$.
	The bottom panel shows the differences between all the observations and the best fit that we obtained with a double power law parameterization (i.e the \textit{dash line} of the upper graphic).
	The solid curves are the best estimate, maximum and minimum values allowed by the galaxy counts at 15,24,70,170 and 850\,$\mu$m in Le Borgne et al. (A\&A, submitted).
}
\end{figure}
\begin{figure}
	\includegraphics[width=9.5cm]{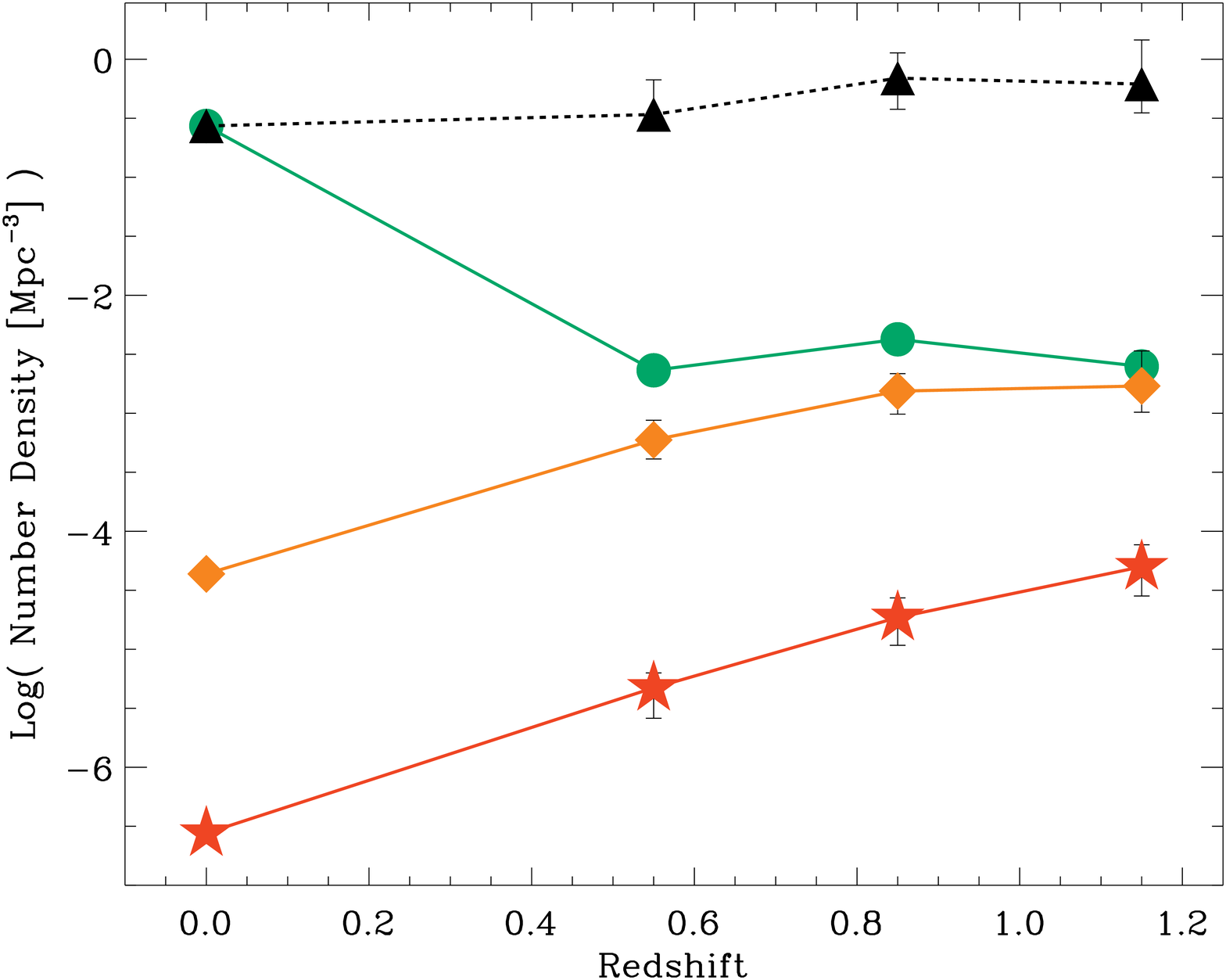}
	\caption{\label{fig:nombre evol}Evolution of the comoving number density up to $z\thicksim1.3$ of normal galaxies (i.e $10^{7}\rm{\,L_{\odot}}<\rm{L_{ir}}<10^{11}\,\rm{L_{\odot}}$; \textit{black filled triangle}), LIRGs (i.e $10^{11}\,\rm{L_{\odot}}<\rm{L_{ir}}<10^{12}\,\rm{L_{\odot}}$; \textit{orange filled diamond}) and, ULIRGs (i.e $10^{12}\,\rm{L_{\odot}}<\rm{L_{ir}}$; \textit{red filled star}).
	The green circles represent the total number of galaxies which are above the detection limit of the surveys presented here, i.e $L_{ir}^{flux\ limit}<\rm{L_{ir}}$.
	The z$\,\thicksim0$ points are taken from \citet{sanders_2003}.
	}
\end{figure}
\indent{
Such decrease of $\phi_{knee}$ was also found by previous studies such as \citet{perez_2005} for the rest-frame 12\,$\mu$m LF and \citet{caputi_2007} for the rest-frame 8\,$\mu$m and the total infrared LF.
Nevertheless this decrease has to be taken with caution since incompleteness in the faint luminosity bins could mimic this evolution \citep{perez_2005}.
\\}
\indent{
Finally, we also notice that the redshift evolution of the rest-frame 35 $\mu$m and total IR LF are in agreement, i.e a pure luminosity evolution proportional to $(1+z)^{3.6\pm0.5}$ and $(1+z)^{3.6\pm0.4}$ for the rest-frame 35 $\mu$m and total IR LF respectively.
This consistency arises from the quasi-linearity between $L_{35}$ and $L_{IR}$.
\\}
\subsection{Comparison with previous work}
\indent{
As revealed in Table \ref{tab:evolution} the global evolution of the total infrared LF inferred in this study is broadly consistent with previous studies \citep{lefloch_2005,caputi_2007,huynh_2007}.
However slight differences appear between in the LIRG and ULIRG regimes.
In Figure \ref{fig:LF IR} we compare the LF inferred in our study with the work of \citet{lefloch_2005} and \citet{huynh_2007}.
\\}
\indent{
We notice that the LF inferred by \citet{lefloch_2005} is consistent with our values at low luminosities.
However, above log$(L_{IR}/L_{\odot})>$11.5, \citet{lefloch_2005} obtained higher values, with a discrepancy increasing with luminosity.
These differences, which affect the relative contributions of LIRGs and ULIRGs to the total infrared luminosity density, might be explained by the fact that we removed AGNs from our catalogs while \citet{lefloch_2005} did not.
However we checked that keeping AGNs to derive the total infrared LF only reduces part of the discrepancy.
Hence the main explanation of these differences is the method used to convert monochromatic luminosities into total infrared luminosities.
\citet{lefloch_2005} used a combination of SED libraries to derive the total infrared luminosity from 24 $\mu$m observations while we used the CE01 library and either direct or stacked 70 $\mu$m observations for the bright and faint end slopes of the LF, respectively.
The latter should provide more robust estimates of $L_{IR}$ since the rest-frame 35 $\mu$m luminosity is produced by similar carriers as the rest-frame 24 $\mu$m one which was proven to be a robust SFR indicator by \citet{calzetti_2007}.
\\}
\indent{
We notice on Figure \ref{fig:LF IR} that our LF is, within the error bars, in agreement with \citet{huynh_2007}.
Nevertheless we also notice that at low luminosities, their values are slightly lower than ours.
We interpret this discrepancy as a difference in the treatment of incompleteness.
Our multi-wavelengths catalogs being computed with a new extraction technique based on prior positions, we are able to extract blended sources and hence reduce incompleteness corrections.
We conclude that at low luminosities the LFs measured in this study should be more robust than the ones derived by \citet{huynh_2007}. 
\\ \\}
\indent{
The total IR LF derived here is noticeably different from the one previously obtained by \citet{caputi_2007} at $z\thicksim1$ (see Figure \ref{fig:lir caputi}) who used 24 $\mu$m images in the CDFS.
There are two possible explanations for this difference: it could either result from cosmic variance or from the method that they used to derive $L_{IR}$ and which does not take into account 70 $\mu$m observations as in our study.
To exclude the first possibility, we computed the total IR LF from our EGS 24 $\mu$m catalog.
Depending on the SED library that we used, we either find an IR LF overlapping ours (using the CE01 SEDs) or matching the \citet{caputi_2007} one (using the $\nu L_{\nu}^{24 \mu \rm m}$-$L_{\rm IR}$ correlation of Bavouzet et al. 2008 as in Caputi et al. 2007).
This shows that the difference between the two studies does not come from cosmic variance but rather from the choice of the SED library.
As we showed earlier, the CE01 SED turned out to fit better the 24 versus 70 $\mu$m correlation down to the faintest luminosities using stacking.
Since the 70 $\mu$m band provides a more accurate estimate of the total infrared luminosity than the 24 $\mu$m one and since the LDP library does not reproduce the observed 24 -- 70 $\mu$m correlation at $z\simeq 1$ (see Figure \ref{fig:stacking}), we conclude that our inferred LFs should be more robust.
\\}
\section{Discussion}
\label{sec: discussion}
\indent{
The redshift  evolution of the comoving number density of LIRGs and ULIRGs is derived by integrating the infrared LF in each redshift bin, $0.4<z<0.7$, $0.7<z<1.0$ and $1.0<z<1.3$.
The number density of LIRGs and ULIRGs appear to be 40 and 100 times larger at $z\thicksim1$ than in the local Universe, respectively (see Figure~\ref{fig:nombre evol}).
Hence we confirm the rapid evolution of these extremely luminous dusty galaxies over the last 8 billion years.
The evolution is at a slightly more moderate pace than previous studies since \citet{lefloch_2005} and \citet{chary_2001} found a factor 70 for LIRGs.
This correction towards a larger contribution of the remaining lower luminosity galaxies, that we name here "normal" galaxies, comes from the faint part of the LF that we were able to constrain through our stacking analysis.
It must be noted that at the highest redshifts, we are only sensitive to the LIRG and ULIRG population
Thus the extrapolation to fainter luminosities (illustrated by the green filled circles in Fig.~\ref{fig:nombre evol}) assumes the same faint end slope  for the LF seen at lower redshifts and could be uncertain.
Finally, we note that the number density of ULIRGs at $z\thicksim1.15$ is nearly equal to that of LIRGs at $z\thicksim0$. In order to detect 10 ULIRGs at $z\sim$1$\pm$0.2 (892 Mpc$^3$/arcmin$^2$), one would need to cover more than 3 GOODS fields.
 \\}
 \begin{table*}
\caption{\label{tab:evolution}Parameters leading the total infrared LF evolution}
\centering
\begin{tabular}{lccccc}
\hline \hline
{Authors} &Functional Form$^{\,\mathrm{a}}$ &{Wavelength}&  &{\rm{$\alpha_{L}$}} &{\rm{$\alpha_{\phi}$}} \\
\hline
This study & DPL & 24 \& 70 $\mu$m & & $3.6\pm0.4$ & $-0.8\pm0.6$ \\
\citet{lefloch_2005} & DE & 24 $\mu$m &  & $3.2^{+0.7}_{-0.2}$ & $0.7^{+0.2}_{-0.6}$ \\
\citet{caputi_2007} & DE  & 24 $\mu$m & $z<1$ & $3.5\pm 0.4 $ & $-0.7\pm0.1$ \\
& & & $z>1$&$2.2\pm0.5$ & $-3.9\pm1.0$ \\
\citet{huynh_2007} & DPL  & 70 $\mu$m &  &$2.78^{+0.34}_{-0.32}$ & $0.0$ \\
\citet{hopkins_2004} & $\cdots$ & SFR compilation &  &$2.7\pm0.6$ & $0.15\pm0.6$\\
\hline
\end{tabular}
\begin{list}{}{}
\item[$^{\mathrm{a}}$] DPL: Double Power Law; DE: Double Exponential 
\end{list}
\end{table*}
 \begin{figure*}
 \begin{center}
	\includegraphics[width=13.cm]{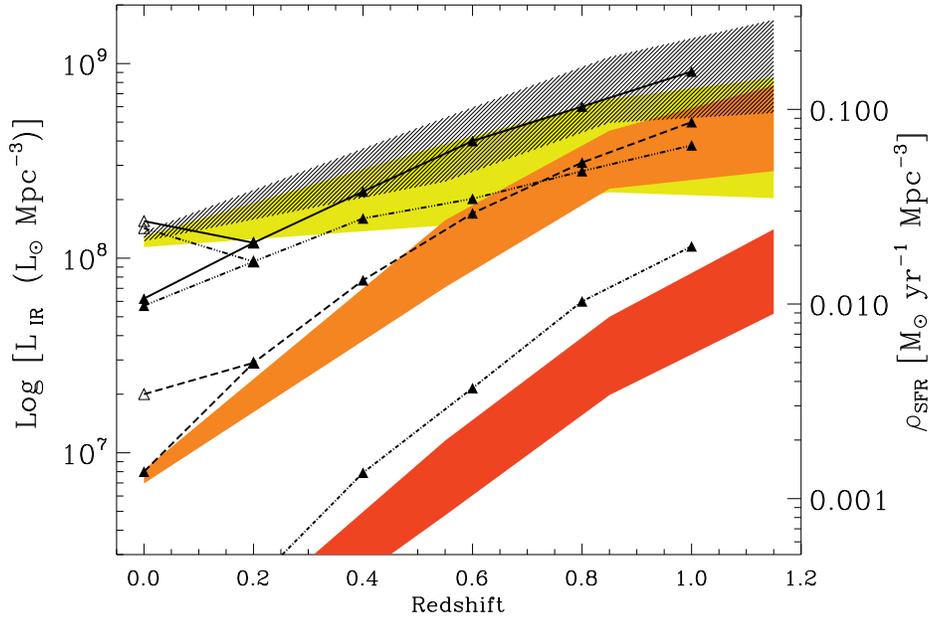}
	\caption{\label{fig:densite evol}Evolution of the comoving IR energy density up to $z\thicksim1.3$ (\textit{striped area}) and the relative contribution of normal galaxies (i.e $7<\rm{log(L_{ir})}<11$; \textit{yellow filled area}), LIRGs (i.e $11<\rm{log(L_{ir})}<12$; \textit{orange filled area}) and, ULIRGs (i.e $12<\rm{log(L_{ir})}$; \textit{red filled area}).
	The areas were defined using all the solutions compatible within 1 $\sigma$ with the total infrared LF.
	Black triangles represent the results obtained by \citet{lefloch_2005} for the global evolution of the comoving energy density (solid line) and the relative contribution of normal galaxies (\textit{triple dot dash line}), LIRGs (\textit{dashed line}) and, ULIRGs (\textit{dot dash line}).
	At $z\thicksim0$ empty black triangles represent results of \citet{lefloch_2005} corrected by a factor 2.5 (see text).
	}
\end{center}
\end{figure*}
\begin{figure*}
 \begin{center}
	\includegraphics[width=13.cm]{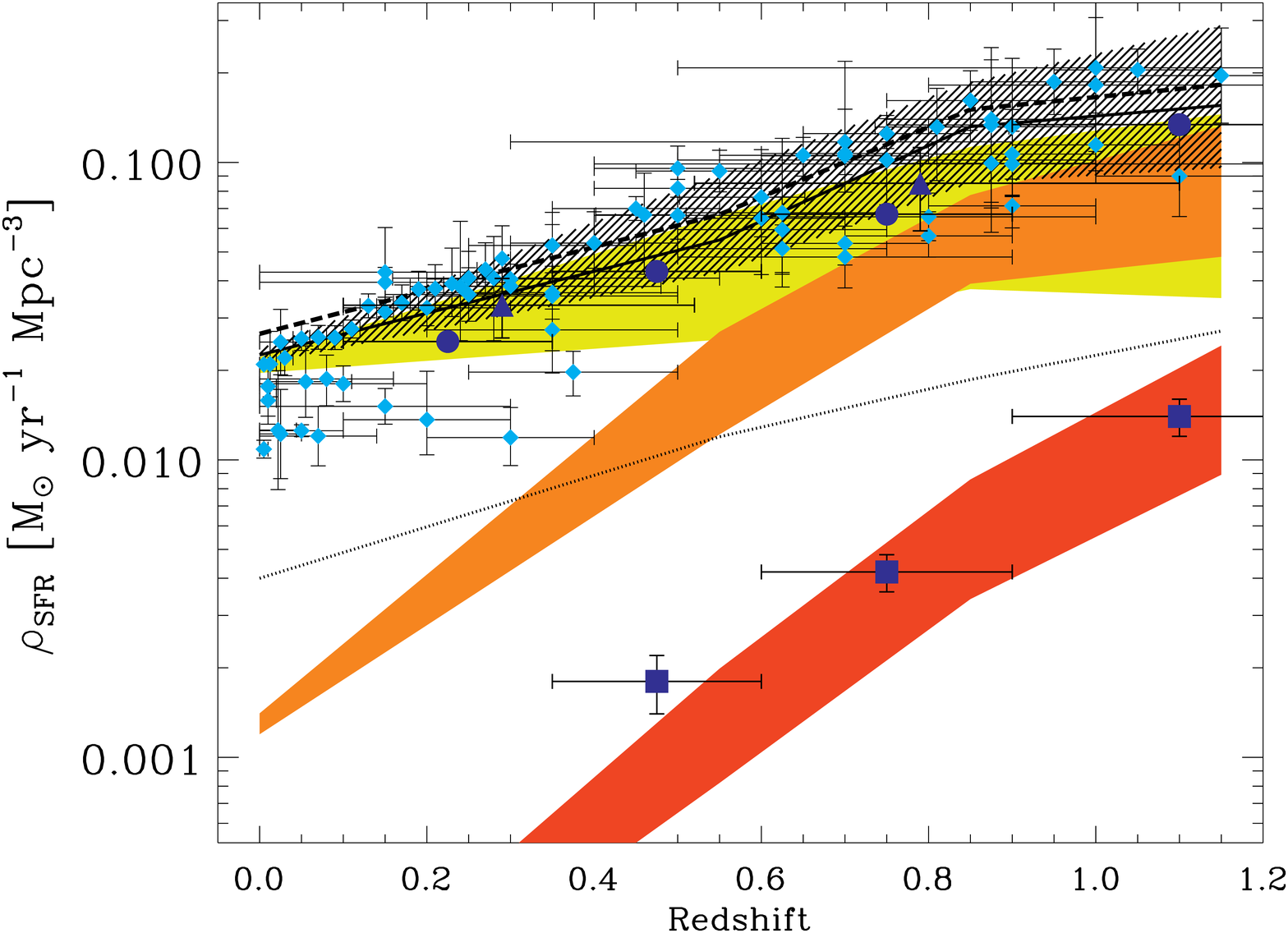}
	\caption{\label{fig:densite evol 2}Evolution of the comoving star formation rate density up to $z\thicksim1.3$ (\textit{striped area}) and the relative contribution of normal galaxies (i.e $7<\rm{log(L_{ir})}<11$; \textit{yellow filled area}), LIRGs (i.e $11<\rm{log(L_{ir})}<12$; \textit{orange filled area}) and, ULIRGs (i.e $12<\rm{log(L_{ir})}$; \textit{red filled area}).
	The areas were defined using all the solutions compatible with the total infrared LF within 1 $\sigma$.
	The solid line represents the best fit to the total SFR density.
	The dottes line represents the SFR measured using the UV light not corrected from dust extinction.
	The dashed line represents the total SFR density defined as the sum of the SFR density estimated by the infrared and of the SFR density obtained from the UV light uncorrected of dust extinction.
	Light blue diamonds are taken from \citet{hopkins_2006} and represent the SFR densities estimates using various estimators.
	Dark blue triangles represent the SFR density estimated by \citet{seymour_2008} using deep radio observations.
	Dark blue circles represent the SFR density estimated by \citet{smolcic_2009} using deep 20 cm observations and dark blue squares represent the relative contribution of ULIRGs to this SFR density.
	}
\end{center}
\end{figure*}
 \indent{
In Figure \ref{fig:densite evol}, we compare the comoving IR luminosity density (or equivalently SFR density due to obscured star formation, using Eq.~\ref{eq:SFR_Lir})  produced by normal galaxies, LIRGs and ULIRGs to that obtained by \citet{lefloch_2005}.
Note that X-ray AGN have already been removed from the LF.
\\}
\indent{ 
We also notice that at $z\thicksim0$ values quoted in \citet{lefloch_2005} need to be corrected by a factor 2.5 (\textit{empty triangles} of Figure \ref{fig:densite evol}).
The total infrared LF taken as their local reference was derived in \citet{sanders_2003} by bins of magnitude but used in \citet{lefloch_2005} as if it was derived by bins of luminosity.
This error does not change the LFs derived at higher redshift by \citet{lefloch_2005} but could have affected their estimates of $\rm{\alpha_{d}}$ and $\rm{\alpha_{L}}$ and hence their estimates of the evolution of the infrared luminosity density.
The discrepancy found between the $\rm{\alpha_{d}}$ derived in this study and the one derived in \citet{lefloch_2005} can be partly due to this error since starting from an erroneous local LF would yield to an overestimate of $\rm{\alpha_{d}}$.
\\}
\indent{
As observed in Figure \ref{fig:densite evol} we confirm the transition between a regime dominated by normal galaxies at low redshift and by LIRGs at high redshift.
The transition occurred at $z\thicksim0.9$ compared to $z\thicksim0.7$ derived by \citet{lefloch_2005}.
While LIRGs and ULIRGs altogether produce less than 2\,$\%$ of the present-day bolometric luminosity density of galaxies, they produced half of it 8 billion years ago.
Indeed, the contribution of unobscured UV light to the comoving SFR density was lower than that of LIRGs/ULIRGs by a factor of about $\sim$3 (see Fig.~\ref{fig:densite evol 2}).
Although the luminosity density of ULIRGs was much larger at $z\thicksim1$ than today, their contribution to the total IR energy density was still rather low (i.e $\thicksim 10\%$).
We find it to be lower than \citet{lefloch_2005} by a factor of $\thicksim2$ which may partly be explained by the fact that unlike \citet{lefloch_2005} we have removed the contribution of X-ray AGN which becomes large among ULIRGs.
Furthermore, by using the observed 70 $\mu$m data instead of extrapolating from 24\,$\mu$m, we have reduced the uncertainty in the estimates of the bolometric luminosity.
\\}
\indent{
In order to get a complete census on the SFR history we need to take into account the contribution of unobscured UV light.
The unobscured SFR density (dot line in Figure \ref{fig:densite evol 2}) is taken from a GALEX study \citep{schiminovich_2005} which yield an evolution following a (1+z)$^{2.5}$ law and a local star formation density of $\rho_{\rm SFR}$($z\sim$0)=$4\times10^{-3}$ $\rm{M_{\odot}yr^{-1}Mpc^{-3}}$.
The total SFR density (dashed line in Figure \ref{fig:densite evol 2}) was then defined as the sum of the unobscured SFR density traced by the direct UV light and the dusty SFR density traced by the infrared emission.
We find that the relative contribution of unobscured UV light to the cosmic SFR density evolves nearly in parallel with the total one and accounts for $\sim$20\,\% of the local SFR density and $\sim$20\,\% of the $\rho_{\rm SFR}$($z\sim$1.15)=0.15 $\rm{M_{\odot}yr^{-1}Mpc^{-3}}$.
Globally, the cosmic star-formation history that we derived using data at 24 and 70 $\mu$m is consistent with the combination of indicators, either unobscured or corrected for dust extinction, as compiled by \citet{hopkins_2006}.
We also notice a very good agreement between the cosmic star-formation history derived in our work and the ones derived by \citet{seymour_2008} and \citet{smolcic_2009} using deep radio observations.
Especially, it is very encouraging to see that the star formation history for ULIRGs obtained using deep radio data is in very good agreement with our independent measurements.
\\}
\section{Conclusion}
\indent{
We make use of the deep 24\,$\mu$m and 70\,$\mu$m surveys in the Great Observatories Origins Deep Survey (GOODS) and Extended Groth Strip (EGS) fields to characterize the evolution of the comoving star-formation rate density at $0.4<z<1.3$ which corresponds to 2/3 of cosmic time.
The data span a total area of $\sim$800 arcmin$^{2}$ and a comoving volume of 1.4$\times$10$^{6}$\,Mpc$^{3}$.
Thanks to the wealth of multiwavelength data, $\sim$80\% of the sources are associated with either a photometric or spectroscopic redshift.
\\}
\indent{
The exquisite depth of the data results in a high surface density of sources. In order to alleviate the effects
of confusion, we have developed a technique which uses the prior positional information of sources at shorter
wavelengths to extract the flux density of sources at 24\,$\mu$m and 70\,$\mu$m. We have undertaken extensive
simulations to quantify flux errors and biases which might be associated with this technique. 
The relative depth of the 24 and 70 $\mu$m survey results in only 7\% of 24 $\mu$m sources being detected at 70 $\mu$m.
\\}
\indent{
By associating the redshifts with the sources in the mid- and far-infrared catalogs, we are able to measure the rest-frame 15\,$\mu$m and 35\,$\mu$m luminosities of galaxies out to $z\sim1.3$.
We use a stacking procedure to measure the evolution of the 24 to 70 $\mu$m luminosity ratio for sources which are undetected at 70 $\mu$m.
We find that the average observed 24\,$\mu$m and 70\,$\mu$m properties of both the detected and stacked sources are in excellent agreement with correlations derived from local galaxy samples.
However, the 24/70 micron flux ratios of individual galaxies do show a significant scatter of 0.2 dex.
\\}
\indent{
By complementing the direct detections with the stacking analysis, we are able to measure a 15\,$\mu$m, 35\,$\mu$m and total infrared luminosity function for star forming galaxies between $0.4<z<1.3$. We measure both the bright end and faint end slope of the luminosity functions, which span a factor of $\sim$100 in luminosity ($\rm{10^{11}\,L_{\odot}< L_{IR}< 10^{13}\,L_{\odot}}$), a substantial
improvement over previous works. We find no
evidence for a change in slope of the double power-law used to characterize the local IRLF of galaxies.
However, we confirm the strong evolution in the number density of infrared luminous galaxies between $0<z<1$.
\\}
\indent{
LIRGs and ULIRGs evolve at about the same rate over this redshift, having increased by a factor of $\sim$40 and $\sim$100 in number
density, respectively.
LIRGs contribute about $\sim$50\% to the comoving star-formation rate density at $z\sim1$, comparable to the contribution of lower luminosity star-forming galaxies.
The infrared luminosity density is a factor of five higher than the UV luminosity density at $z\sim1$ confirming that the bulk of the star-formation is dust obscured.
\\}
\acknowledgements{
This work is based on observations made with the Spitzer Space Telescope, which is operated by the Jet Propulsion Laboratory, California Institute of Technology under a contract with NASA. 
Support for this work was provided by NASA through an award issued by JPL/Caltech.
This paper makes use of photometric redshifts produced jointly by Terapix and VVDS teams.
Funding for the DEEP2 survey has been provided by NSF grants AST95-09298, AST-0071048, AST-0071198, AST-0507428, and AST-0507483 as well as NASA LTSA grant NNG04GC89G.
Some of the data presented herein were obtained at the W. M. Keck Observatory, which is operated as a scientific partnership among the California Institute of Technology, the University of California and the National Aeronautics and Space Administration. The Observatory was made possible by the generous financial support of the W. M. Keck Foundation. The DEEP2 team and Keck Observatory acknowledge the very significant cultural role and reverence that the summit of Mauna Kea has always had within the indigenous Hawaiian community and appreciate the opportunity to conduct observations from this mountain. D.Le Borgne and D.Elbaz wish to thank the Centre National d'Etudes Spatiales (CNES) for their support.
}
 \bibliographystyle{aa}

\end{document}